\documentclass[a4paper,12pt]{article}
\pdfoutput=1

\usepackage[normalem]{ulem}

\usepackage[utf8]{inputenc}
\usepackage{amsmath,amssymb,amsthm}
\usepackage{dsfont}%for unit operator
\usepackage{physics,mathtools}
\usepackage{tikz,float}
\usepackage{booktabs}%For tables
\usepackage[hidelinks]{hyperref} % Enables clickable links in TOC
\numberwithin{equation}{section}  % make eq labels (sec.num)
\usepackage{cite} 
\usepackage{subcaption} % modern package for subfigures
\usepackage{caption}    % for consistent caption formatting
\makeatletter
\def\@fpheader{\vspace{0.1mm}}
\makeatother
\newcommand{\bw}{\begin{widetext}}
\newcommand{\ew}{\end{widetext}}
\newcommand{\bea}{\begin{eqnarray}}
\newcommand{\eea}{\end{eqnarray}}
\newcommand{\be}{\begin{equation}}
\newcommand{\ee}{\end{equation}}
\newcommand{\bca}{\begin{cases}}
\newcommand{\eca}{\end{cases}}

\def\b{{\beta}}

\newcommand\lam{\lambda}
\newcommand\Lam{\Lambda}

\newcommand\Ga{{\ensuremath{{\Gamma}}}}
\newcommand\de{{\ensuremath{{\delta}}}}
\newcommand\De{{\ensuremath{{\Delta}}}}

\newcommand\ov{\over}

\def\le{\left}
\def\ri{\right}
\newcommand{\ben}{\begin{enumerate}}
\newcommand{\een}{\end{enumerate}}
\newcommand{\coo}[1]{{\mathcal{O}\left(#1\right)}}

\def\cL{{\cal L}}

\definecolor{reed}{rgb}{0.6, 0.0, 0.0}

\usetikzlibrary{patterns}
\makeatletter
\let\over\@@over
\makeatother

\def\qt{{\tilde q}}

\hypersetup{
  colorlinks   = true, %Colours links instead of  boxes
  linkcolor={blue!75!green},
    citecolor={red!90!black},
    urlcolor={blue!80!black}
}

\usepackage{tocloft}

\begin{document}

\begin{titlepage}
  \thispagestyle{empty}
%\preprint{MIT-CTP/5957}
\hfill\text{MIT-CTP/5957}

\begin{center}

{\LARGE \scshape{Fooling the Censor:\\
\vspace{0.2cm}
\noindent
\makebox[\dimexpr\textwidth+2cm][l]{%
\parbox{\dimexpr\textwidth+2cm}{%
Going beyond inner horizons with the OPE
}}}}

\vskip1.1cm 

\centerline{\textbf{ 
   {Nejc \v{C}eplak}${}^1$,
   {Hong Liu}${}^{2}$,
   {Andrei Parnachev}${}^1$,
   {and Samuel Valach}${}^3$
}}

\vskip0.65cm
 
${}^1${\footnotesize{\noindent{\em School of Mathematics and Hamilton Mathematics Institute, Trinity College, Dublin 2, Ireland
}}}
 
\vskip-0.07cm
 
${}^2${\footnotesize{\noindent{\em MIT Center for Theoretical Physics---a Leinweber Institute, Massachusetts Institute of Technology, Cambridge, MA 02139, USA
}}}

\vskip-0.07cm

${}^3${\footnotesize{\noindent{\em Faculty of Mathematics and Physics, University of Ljubljana, SI-1000, Ljubljana, Slovenia
}}}

\end{center}

\vskip0.35cm
%

% Emails
\centerline{\footnotesize\upshape\ttfamily  
ceplakn@tcd.ie, hong\_liu@mit.edu, parnachev@maths.tcd.ie, samuel.valach@fmf.uni-lj.si}

\bigskip

\begin{abstract}
  \noindent 

The analytic structure of holographic correlation functions at finite temperature contains information about curvature singularities of black holes in AdS.
We compute the Operator Product Expansion (OPE) coefficients of the holographic two-point function of scalar operators at finite temperature and finite chemical potential. 
We show that the stress-tensor and current (T+J) sector of the OPE contains a singularity in the complex time plane at a location that can be identified with the time-shift of a bouncing geodesic in the charged black hole geometry: The geodesic starts at a boundary of a charged black hole in AdS, 
bounces off the timelike singularity, before returning to a different asymptotic boundary on the same side of the Penrose diagram.
We show that the singularity in the   T+J sector is smooth across the point where black hole becomes extremal, indicating that the analytic properties of holographic correlators could potentially probe naked singularities.

\end{abstract}

\begin{figure}[!htb]
    \centering
    \vspace{-0.2cm} % adjust if you want it closer to the abstract
    \makebox[\textwidth][c]{%
        \includegraphics[scale=0.30]{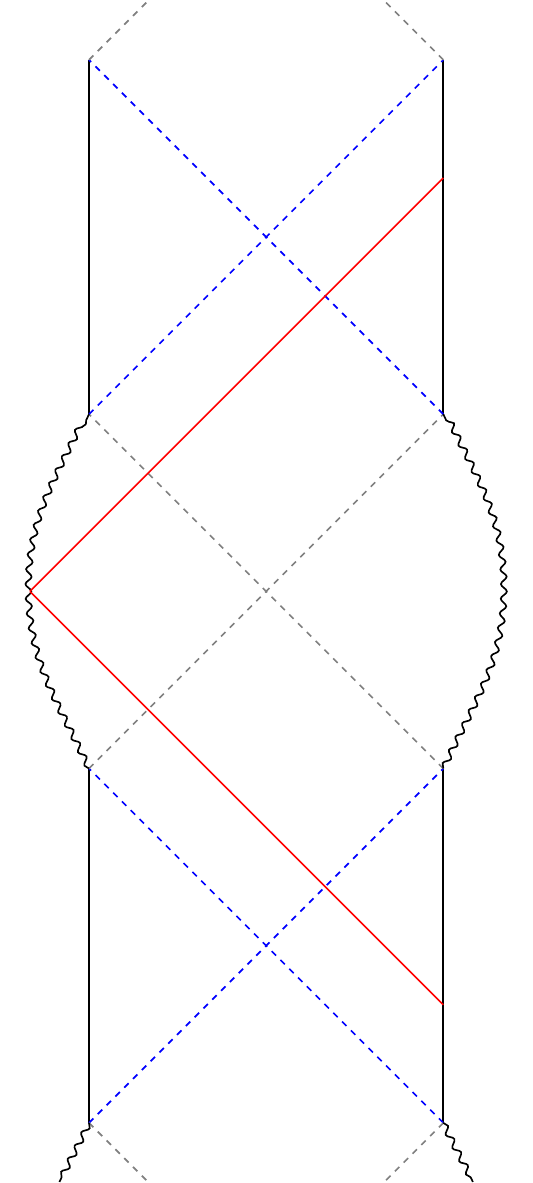}%
    }
    \vspace{-4cm} % adjust bottom spacing if needed
\end{figure}

\end{titlepage}

\setcounter{tocdepth}{2}
{\small \tableofcontents}

\clearpage

%%%%%%%%%%%%%%%%%%%%%%%%%%%%%%%%%%%%%%%%%%%%%%%%%%%%%
%%%%%%%%%%%%%%%%%%%%%%%%%%%%%%%%%%%%%%%%%%%%%%%%%%%%%

\section{Introduction and summary}
\label{sec:introduction}

In general relativity, curvature singularities, such as those inside black holes, mark loci where the classical description fails and a more fundamental theory of quantum gravity takes over.
The weak and strong cosmic censorship conjectures assert, respectively, that naked singularities do not form and that no observer can cross a Cauchy horizon, for generic and physically reasonable initial data~\cite{Penrose:1969pc, Penrose:1980ge}.

\begin{figure}[t]
    \centering
    \includegraphics[scale=1]{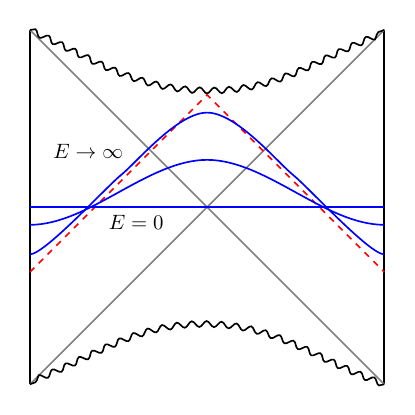}
    \caption{The Penrose diagram of an eternal Schwarzschild-AdS black hole in four dimensions or higher. The curvature singularity is bent inwards compared to the asymptotic boundary.
    Spacelike geodesics (blue) with endpoints on different boundaries probe the interior region. 
    As their energy, $E$ is increased, the geodesics tend to a null geodesic that reflects off the black hole singularity (red-dashed).}
    \label{fig:Penrose}
\end{figure}

The AdS/CFT duality~\cite{Maldacena:1997re, Gubser:1998bc, Witten:1998qj}, which in principle encodes all aspects of black hole physics in AdS, provides a fruitful framework for studying the black hole interior and singularities; see, for example,~\cite{Louko:2000tp,Kraus:2002iv,Levi:2003cx,Fidkowski:2003nf,Brecher:2004gn,Balasubramanian:2004zu,Festuccia:2005pi,Festuccia:2006sa,Amado:2008hw,Hartman:2013qma,LiuSuh13a,LiuSuh13b,Grinberg:2020fdj,Rodriguez-Gomez:2021pfh,Leutheusser:2021frk,deBoer:2022zps,David:2022nfn,Horowitz:2023ury, Kolanowski:2023hvh, Parisini:2023nbd,Dodelson:2023nnr,Singhi:2024sdr,Arean:2024pzo,Dodelson:2024atp,Cai:2024ltu,Balasubramanian:2019qwk,Ceplak:2024bja,Afkhami-Jeddi:2025wra}.
In particular, signatures of the black hole singularities of an eternal AdS black hole---dual to the thermofield-double state~\cite{Maldacena:2001kr}---probed by bouncing null geodesics (see Figure~\ref{fig:Penrose}) have been identified long ago in the boundary thermal two-point correlation functions~\cite{Fidkowski:2003nf,Festuccia:2005pi}:\footnote{New signature have been  found recently in~\cite{Ceplak:2024bja,Afkhami-Jeddi:2025wra}.}
\begin{align}
\label{eq:CorrelatorIntro}
G(\tau) = \left\langle \phi(\tau) \phi(0) \right\rangle_{\beta} \,.
\end{align}
Here $\phi$ is a scalar operator of conformal dimension $\Delta$, and $\beta$ is the inverse temperature.
The variable $\tau$ is the complexified Euclidean time,
\begin{align}
\tau = t_E + i t_L\, ,
\end{align}
whose real and imaginary parts correspond to the Euclidean and Lorentzian time separations, respectively.\footnote{Throughout, we set the spatial separation $\vec{x} = 0$.} 

Associated with a bouncing null geodesic is a time shift
\begin{align}
\label{eq:TimeShiftIntro}
    \tau_c^{\pm} = \pm i\,\frac{\beta\,e^{\mp i \frac{i\pi}{d}}}{2\sin\frac{\pi}{d}} = \frac{\beta}{2} \pm i \frac{\beta}{2}\,\cot\frac{\pi}{d}\,.
\end{align}
giving the time difference between the two boundary points  connected by the geodesic. 
The real part arises from traversing the horizons of the black hole, while the imaginary part reflects the fact that the singularity in the Schwarzschild-AdS geometry is ``bent inward'' relative to the asymptotic boundary~\cite{Fidkowski:2003nf}.
The sign $+$ (-) corresponds to bouncing off the future (past) singularity. 
In $G(\tau)$, the quantity $\tau_c$ either appears as a singularity after suitable analytic continuation in the large-$\Delta$ limit~\cite{Fidkowski:2003nf}, or governs the asymptotic behavior of the correlation function along the imaginary frequency axis~\cite{Festuccia:2005pi}.

Recently, significant new light on the CFT origin of these signatures has been shed by examining the Operator Product Expansion (OPE) structure of~\eqref{eq:CorrelatorIntro}~\cite{Ceplak:2024bja}. More explicitly, by performing the OPE of $\phi(\tau)\phi(0)$, one can express $G(\tau)$ as
\begin{align}\label{e.tcbdecoomp}
    G(\tau) = \frac{1}{\tau^{2\Delta}}\,\sum_{n}\,C_n\,v_n\left(\frac{\tau}{\beta}\right)^{\Delta_n}, \qquad {\left\langle\mathcal{O}_n\right\rangle_{\beta}} = \frac{v_n}{\beta^{\Delta_n}}\,,
\end{align}
where the sum runs over all operators ${\cal O}_n$ (with conformal dimension $\Delta_n$) appearing in the OPE that have a non-trivial thermal one-point function, with $C_n$ being the associated OPE coefficients.

In the regime dual to Einstein gravity in the bulk, $G(\tau)$ can be decomposed into two distinct contributions:
\begin{align} \label{decom}
G(\tau) = G_{\rm T}(\tau) + G_{[\phi\phi]}(\tau)\,,
\end{align}
where $G_{\rm T}(\tau)$---the {\it stress-tensor sector} contribution---includes the exchanges of the identity, the stress tensor, and its composites, while $G_{[\phi\phi]}(\tau)$, the \textit{double-trace sector} contribution, includes the contributions from the exchanges of double-trace operators built from the insertion fields $\phi$, schematically of the form $\phi(\partial^2)^n\partial_{\mu_1}\cdots\partial_{\mu_l}\phi$.%
\footnote{The double-trace sector also includes composites of $\phi$ fields and the stress-tensor fields. }

It was found in~\cite{Ceplak:2024bja} that $G_{\rm T}$, for general $\Delta$, exhibits a singularity at~\eqref{eq:TimeShiftIntro}:
\begin{align}
\label{eq:SingularityIntro}
G_{\rm T}(\tau \to \tau_c) \sim \frac{1}{\left(\tau_c - \tau\right)^{2\Delta - \frac{d}{2}}}\,, 
\end{align}
which provides a new signature of the bouncing geodesic and can further be used to give a boundary explanation of the signatures identified in~\cite{Fidkowski:2003nf,Festuccia:2005pi}. Note that the individual sectors $G_{\rm T}$ and $G_{[\phi\phi]}(\tau)$ are much less constrained than the full correlator:  they do not have to be analytic in the strip ${\rm Re} \, \tau \in (0, \b)$ and do not need to  
satisfy the KMS condition separately~\cite{Ceplak:2024bja}\footnote{See \cite{El-Showk:2011yvt,Iliesiu:2018fao,Iliesiu:2018zlz,Alday:2020eua,Marchetto:2023xap,Barrat:2025wbi,Buric:2025anb,Barrat:2025nvu,Buric:2025fye,Niarchos:2025cdg} for recent work devoted to the KMS conditions in CFTs.}.

To derive~\eqref{eq:SingularityIntro}, the scheme developed in \cite{Fitzpatrick:2019zqz} (see also Appendix~\ref{app:OPE}) was used to calculate thermal OPE coefficients associated to the stress-tensor exchanges\footnote{Similar results can be obtained using Fourier transform methods \cite{Parisini:2023nbd,Afkhami-Jeddi:2025wra}.
Also for a selection of papers where the OPE decomposition of holographic thermal two-point functions was investigated see
\cite{Karlsson:2019qfi,Li:2019tpf,Kulaxizi:2019tkd,Fitzpatrick:2019efk,Karlsson:2019dbd,Li:2019zba,Karlsson:2020ghx,Li:2020dqm,Fitzpatrick:2020yjb,Karlsson:2022osn,Dodelson:2022yvn,Parisini:2022wkb,Dodelson:2023vrw,Haehl:2025ehf,Bajc:2025jjv,Barrat:2025twb
}. 
Once the stress-tensor sector is known,  one can compute the thermal coefficients of the double traces
by imposing the KMS condition \cite{Buric:2025anb, Buric:2025fye}.}.
This involves solving the bulk equation of motion of a minimally coupled scalar in the Euclidean black hole background in a near-boundary expansion. From the bulk perspective, the upshot is that the signatures of the black hole singularity are encoded in data that depend only on the asymptotic structure of the spacetime and carry no direct information about the black hole interior.

\begin{figure}[t]
    \centering
    \includegraphics[scale=0.5]{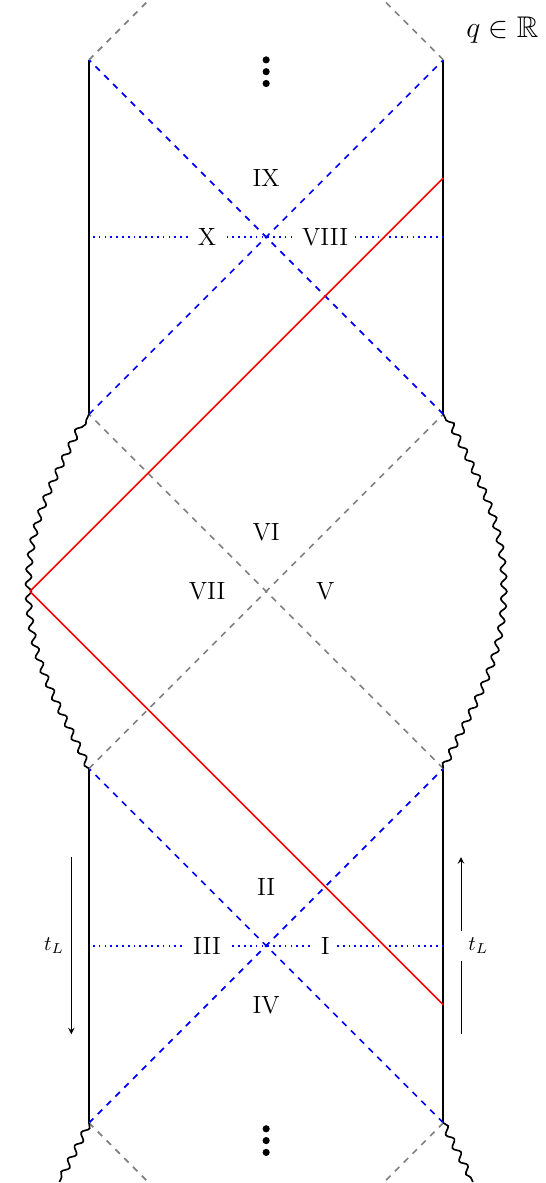}
    \caption{The Penrose diagram for a charged black hole with real charge $q$. The outer horizons are depicted in blue dashed lines, while the inner horizons are in dashed gray. The timelike singularity is bent out compared to the asymptotic boundary~\cite{Brecher:2004gn}. 
    We will argue that the boundary correlation functions contain a singularity which coincides with the time shift of a bouncing geodesics that starts at the asymptotic boundary of region I, reflects off the singularity in region VII and reaches the asymptotic boundary on the same side of the diagram, but at a different asymptotic boundary -- that of region VIII.}
    \label{fig:PDRealQ}
\end{figure}

In this paper, we explore the boundary signatures of the singularities of an electrically charged black brane, which is dual to a state at finite temperature and finite charge~\cite{Chamblin:1999hg,Chamblin:1999tk}.
From the bulk point of view, adding any amount of charge (that backreacts on the geometry) drastically alters the spacetime causal structure. As illustrated in the Penrose diagram in Fig.~\ref{fig:PDRealQ}, inner horizons appear in addition to the outer horizon, and the spacelike singularities of the neutral black hole become timelike, hidden behind the inner horizons. Furthermore, there is an upper limit on the amount of charge a black hole can carry
\begin{align}
    \label{eq:ChargeBoundIntro}
    q \leq q_{\rm ext}\,.
\end{align}
Geometry with $q > q_{\rm ext}$ exhibits a naked curvature singularity that is not cloaked by a horizon.

Understanding the possible existence of boundary signatures of the inner horizons and timelike singularities of a charged black hole has long presented a challenge (see, $e.g.$,~\cite{Levi:2003cx,Balasubramanian:2004zu,Brecher:2004gn,Castro:2013kea,Balasubramanian:2019qwk} for earlier discussions).
There are good reasons to believe that regions beyond the inner horizons should be invisible to the boundary theory: the strong cosmic censorship conjecture implies that the region behind the inner horizon is unphysical, as physical perturbations (such as infalling radiation) become infinitely blue-shifted near it, turning the horizon into a curvature singularity and rendering the spacetime inextendible beyond it~\cite{Simpson:1973ua,Poisson:1990eh,Ori:1991zz,Dafermos:2017dbw}.

\begin{figure}[t]
    \centering
    \includegraphics[width=0.95\linewidth]{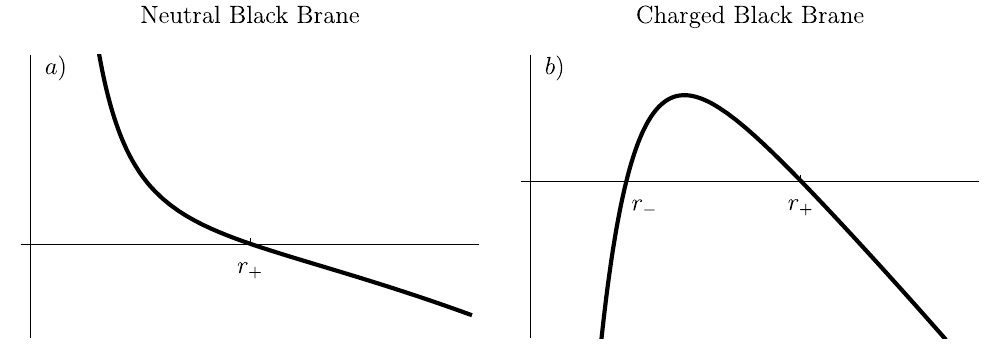}
    \caption{Determining the spacelike geodesics in black hole can be mapped to the scattering of a classical particle in a potential $V(r)$. On the left (a), we depict the potential for a neutral black brane in AdS, where as the energy in increased, the geodesics probe closer and closer to the singularity at the origin, where they approach the bouncing geodesic. 
    On the right (b), we show the potential for the charged black brane, where instead there exists a maximum between the inner ($r_-$) and the outer horizon $r_+$. Spacelike geodesics can only probe to a minimal distance, set by the maximum of the potential and do not tend to a null geodesic.}
    \label{fig:Potentials}
\end{figure}

Extending the bouncing-geodesic construction of~\cite{Fidkowski:2003nf} to charged black holes has also not succeeded in probing regions beyond the inner horizons~\cite{Brecher:2004gn}.
For a neutral black hole, the null bouncing geodesic in Fig.~\ref{fig:Penrose} arises as the infinite-energy limit of two-sided spacelike geodesics, which can in turn be related to the large-dimension limit of $G(\tau)$. This was how the connection between the null geodesic and $G(\tau)$ was established in~\cite{Fidkowski:2003nf}.
The equation for a spacelike geodesic can be mapped to the motion of a classical particle with total energy $E^2$ in a potential
\begin{align}
\label{eq:PotentialIntro}
E^2 = \dot r^2 + V(r)\,, \qquad V(r) = - r^2 f(r)\,,
\end{align}
where $f(r)$ is the blackening factor determining the black hole geometry.
For a neutral black hole, the potential $V(r)$ takes the form shown in the left plot of Fig.~\ref{fig:Potentials}. As the energy $E$ increases, the turning point moves inward, and in the limit $E \to \infty$, it approaches the singularity at $r = 0$, becoming null.
In contrast, in the charged case---shown in the right plot of Fig.~\ref{fig:Potentials}---the potential develops a maximum, implying that at sufficiently large energies a radial spacelike geodesic simply falls into the singularity and does not bounce to the opposite boundary. The geodesics that do bounce ($i.e.$, those with energy below the maximum and reflected by the potential) and reach the opposite boundary can at most approach a finite distance from the black hole singularity, which always lies outside the inner horizon.\footnote{Radial geodesics in charged black holes behave analogously to spacelike geodesics with nonzero angular momentum in neutral black holes~\cite{Fidkowski:2003nf}.}

In this paper, we provide analytic and numerical indications that these difficulties can be circumvented, allowing us to identify signatures of both the timelike singularities and the inner horizons in the boundary theory.
This is achieved, on the one hand, by identifying the appropriate boundary quantity ($i.e.$, a generalization of $G_{\rm T}$ discussed above) that captures such signatures, and on the other hand, by developing a better understanding of the role of the null bouncing geodesics shown in Fig.~\ref{fig:PDRealQ}, as well as of the corresponding~(complex) spacelike geodesics whose infinite-energy limit gives rise to the null bouncing geodesics.
Furthermore, we are able to probe the regime with $q > q_{\rm ext}$---$i.e.$, the geometry containing a naked singularity---using the boundary theory. Our results are not in conflict with either the strong or weak cosmic censorship conjecture, as the charged black brane represents a ``fine-tuned'' geometry (and the corresponding boundary state is likewise ``fine-tuned'').

More explicitly, we consider correlation functions of {\it neutral} scalar operators~\eqref{eq:CorrelatorIntro} in a state at finite temperature and finite charge, and use the OPE to decompose the correlator into two sectors,
\begin{align}
G(\tau) = G_{\rm T+J}(\tau) + G_{[\phi\phi]}(\tau),
\end{align}
where the second term again denotes the double-trace sector contribution, while the first term contains contributions from the exchanges of the identity, the stress tensor, the $U(1)$ current, and their composites---we will refer to it as the T+J sector contribution.
We calculate the OPE coefficients associated with the T+J sector and provide evidence that $G_{\rm T+J}(\tau)$ exhibits a singularity of the form
\begin{align}
\label{eq:MainRes}
G_{\rm T+J}(\tau) \sim \frac{1}{\big(\tau_c(\mu,q) - \tau\big)^{2\Delta - b_{\Delta}(\mu,q)}}\,, 
\end{align}
where $\mu$ and $q$ are the mass and charge of the black hole.\footnote{More precisely, as we will see below, the parameters $\mu$ and $q$ are related to the black hole mass $M$ and charge $Q$ by numerical factors. However, since $\mu$ and $q$ appear as the natural counting parameters in the CFT description, we will, with a slight abuse of terminology, refer to them as the mass and charge. Throughout this paper we will take $\mu >0$, but it will be convenient to consider $q$ as a complex variable. }
Here $b_{\Delta}(\mu,q)$ is an order-$\mathcal{O}(1)$ coefficient, and $\tau_c(\mu,q)$ is the time shift ($i.e.$, boundary time difference of its end points) associated with the null bouncing geodesic shown in Fig.~\ref{fig:PDRealQ}. This shows that the properties of black hole singularities are encoded in the analytic structure of boundary correlation functions, even when the singularities are hidden behind inner horizons, and that the bouncing geodesics connect points on the asymptotic boundaries on the same side of the Penrose diagram.
We have only examined $G_{\rm T+J}(\tau)$ at finite $\Delta$. It is possible that~\eqref{eq:MainRes} could give rise to signatures in the large-$\Delta$ limit of the full correlation function, as in the neutral black hole case, which we leave for future investigation.

Since the bouncing null geodesic in Fig.~\ref{fig:PDRealQ} traverses both the outer and inner horizons, the time shift $\tau_c(\mu,q)$ encodes information about the inner horizon.
For example, for $d=4$, the small-$q$ expansion of $\tau_c(\mu,q)$ takes the form of a double series,
\begin{align} \label{tauX0}
\tau_c(\mu, q) &= \tau_0 + q^2 \tau_1 + q^4 \tau_2 + \cdots \cr
&\quad + (q^2)^{3 / 2} \left( \tau_{3/2} + q^2 \tau_{5/2} + \cdots \right)\,,
\end{align}
where $\tau_0$ is the time shift of the neutral black hole.
The OPE expansion of $G_{\rm T+J}(\tau)$ is a series in integer powers of $q^2$, as we are considering a neutral scalar operator.
Interestingly, the expansion~\eqref{tauX0} contains a nonanalytic fractional-power subseries that arises solely from the contribution of the inner horizon.
We will present an argument that the structure of~\eqref{tauX0}---including the nonanalytic fractional-power subseries associated with the inner horizon---can be understood from the analytic structure of the OPE coefficients.

Another interesting feature of $\tau_c(\mu,q)$ is that it varies smoothly as $q$ crosses the bound~\eqref{eq:ChargeBoundIntro}, and that~\eqref{eq:MainRes} continues to hold in the super-extremal regime, where the bulk geometry contains a naked singularity.
In this regime, the bouncing geodesic shown in Fig.~\ref{fig:PDRealQ} becomes complex.

The rest of the paper is organised as follows.
In section~\ref{sec:GeoPicture} we review the charged black brane solutions. 
In particular, we review their causal structure and potential bouncing null geodesics. 
In Section~\ref{sec:chargedbh} we then calculate the OPE coefficients for the stress-tensor and charge-current sectors of thermal correlation functions of scalar fields. 
We analyse their asymptotic structure and show that they imply a singularity whose location in the complex time plane coincides with the time shift associated with geodesics bouncing off the time-like singularity discussed in the previous section. 
We conclude with a discussion in Section~\ref{sec:Disc}.

In the appendices we summarise some technical details. 
In Appendix~\ref{app:SpaceGeo} we outline the calculation of the proper length of spacelike geodesics in the charged black brane geometries.
The algorithm used to extract the OPE coefficients associated with the T+J sector is summarised in Appendix~\ref{app:OPE}.

\textbf{Note added.} While this paper was in preparation, we became aware of the forthcoming papers  \cite{CERNgroup} and \cite{AliAhmad} which may contain some overlapping results.

%%%%%%%%%%%%%%%%%%%%%%%%%%%%%%%%%%%%%%%%%%%%%%%%%%%%%
%%%%%%%%%%%%%%%%%%%%%%%%%%%%%%%%%%%%%%%%%%%%%%%%%%%%%

\section{Bouncing geodesics in charged black hole}
\label{sec:GeoPicture}

In this section we analyse bouncing geodesics in the charged black brane geometry. 
We reformulate the geodesic computation in terms of contour integrals and compute the time-shift as a function of the charge.
We also discuss the extremal and super-extremal solutions and show how the time-shift can be analytically continued to those regimes.

\subsection{Charge black brane solution}
\label{ssec:qneq0setup}
The electrically charged black brane is a solution to the Einstein-Maxwell theory in a negatively curved spacetime. 
In $(d+1)$ dimensions, the solution can be written as \cite{Chamblin:1999tk, Chamblin:1999hg}
\begin{align}
     \label{eq:ChargedMet1}
    ds^2 = -r^2\,f(r)\,dt^2 + \frac{dr^2}{r^2\,f(r)}+ r^2\,d\vec{x}_{d-1}^{\phantom{.}2}\,,\qquad A = \left(- \frac{1}{c}\,\frac{q}{r^{d-2}}+ \Phi\right)\,dt\,,
\end{align}
where 
\begin{align}
\label{eq:Blackening}
    f(r) = 1- \frac{\mu}{r^{d}}+ \frac{q^2}{r^{2d-2}}\ ,
\end{align}
and $c= \sqrt{2(d-2)/(d-1)}$.
The two parameters $\mu$ and $q$ are related to the mass and the charge of the black brane
and we set the radius of AdS $R_{\rm AdS}=1$ throughout. 
$r$ is the radial coordinate, with the AdS boundary at $r\to \infty$. 
The boundary is $\mathbb{R}_t \times \mathbb{R}^{d-1}$, where the time coordinate is $t$, while the spatial directions are parametrised by $\vec{x} = (x_1, x_2, \ldots, x_{d-1})$
The constant $\Phi$ is the electrostatic difference between the horizon and the asymptotic infinity, which is chosen such that $A_t$ vanishes at the outer horizon. 

We will  focus on $d=4$ in the main text.%
\footnote{The analysis for $d=6$ shows essentially the same features as in $d=4$ presented in the main text. We do not expect the results to vary significantly with dimension, as long as $d>2$, so that the black hole singularity is a true curvature singularity.}
The locations of the outer horizon, $r_+$, and the inner horizon, $r_-$, are determined by the solutions of  $f(r) = 0$, which can be rewritten as
\begin{align}
\label{eq:HorizonEq}
   y^3 - \mu y + q^2= (y- y_+)(y-y_-)(y-y_e)= 0\,,\qquad y \equiv r^2\,.
\end{align}
If $q \in \mathbb{R}$, then one can order the solutions to this equation as $y_+ \geq y_- \geq 0 \geq y_e$, and define 
\begin{align}
\label{eq:Horizons}
    r_+ \equiv \sqrt{y_+}\,,\qquad r_- \equiv  \sqrt{y_-}\,, \qquad r_e \equiv \sqrt{-y_e}\ .
\end{align}
In terms of $q$ and $\mu$,  the locations of the horizons are given implicitly through
\begin{align}
     y_+  + y_- + y_e = 0\,,\qquad y_+\,y_-\,y_e = -q^2\,, \qquad y_+\,y_- + y_+\,y_e + y_-\,y_e = -\mu\,.
\end{align}
It is useful to solve these equations in a small charge expansion ($q^{\frac43} \ll 
\mu$) and obtain
\begin{subequations}
\label{eq:RootExpansions}
    \begin{align}
        r_+ &=
        \sqrt[4]{\mu }-\frac{q^2}{4 \mu ^{5/4}}-\frac{7 q^4}{32
   \mu ^{11/4}}-\frac{39 q^6}{128 \mu ^{17/4}}+\coo{q^8}\,,\\
        \label{rminus}
        r_- &=\sqrt{q^2}\left[
        \frac{1}{\sqrt{\mu }}+\frac{ q^4}{2 \mu
   ^{7/2}}+\coo{q^8}\right]\,,\\
       i  r_e &=i \sqrt[4]{\mu }+\frac{i q^2}{4 \mu ^{5/4}}
       -\frac{7 i
   q^4}{32 \mu ^{11/4}}
   +\frac{39 i q^6}{128 \mu ^{17/4}}
        +\coo{q^8}\,.
    \end{align}
\end{subequations}
It is important to note that the $r_+$ and $r_e$ are analytic in $q^2$ near $q=0$, while the location $r_-$ of the inner horizon is not, with a square root branch point singularity. 

\begin{figure}
    \centering
    \includegraphics[scale=0.5]{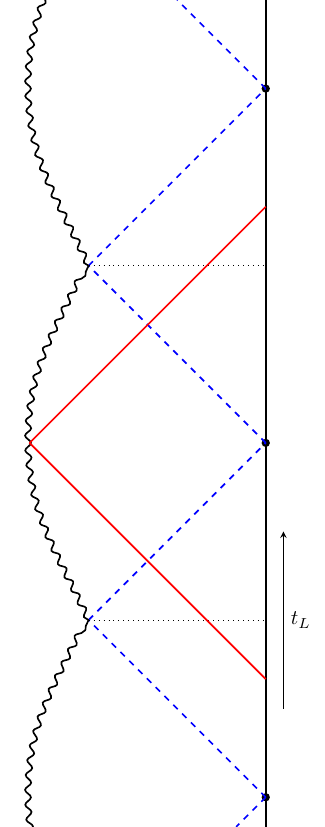}
    \caption{The Penrose diagram for the extremal black hole in AdS. On the left of the diagram there are infinitely many copies of the boundary of AdS, while on the right we find the singularities, which are bent out relative to the boundary. 
    $\tau_c(\mu,q_{\rm ext})$ can be interpreted as the time-shift of a null geodesic (red) starting at one copy of the asymptotic boundary, crossing the degenerate horizon and bouncing off the singularity towards a different copy of the asymptotic boundary. }
    \label{fig:ExtremalPenrose}
\end{figure}

Let us track the causal structure of spacetime as the charge, $q$, of the black hole is increased, while its mass is kept fixed.
At vanishing charge we have the neutral black brane, whose Penrose diagram is depicted in Figure~\ref{fig:Penrose}. 
The geometry has two causally disconnected asymptotic boundaries. 
There is only one event horizon behind which there is a spacelike singularity, which is bend in compared to the asymptotic boundary \cite{Fidkowski:2003nf}.
Adding a small amount of charge, $q \in \mathbb{R}$ drastically changes the causal structure of the black hole, see Figure~\ref{fig:PDRealQ}.  
There is now an inner and an outer horizon, to each we can associate an inverse temperature 
\begin{align}
\label{eq:Temp}
    \beta_\pm = \frac{2\pi\,r_\pm^3}{(r_e^2+ r_\pm^2)(r_+^2-r_-^2)}\ .
\end{align}
$\beta_+$, the inverse temperature at the outer horizon, is taken as the inverse temperature of the dual thermal CFT state.
Unlike in the neutral case, the singularity is timelike and is bent outwards compared to the asymptotic boundaries \cite{Brecher:2004gn}.
The Penrose diagram is infinite -- there are infinitely many copies of the asymptotic regions. 
This structure persists until a critical value of the charge, which in $d=4$ is at
\begin{align}\label{eq:extcond}
    q_{\rm ext} = \sqrt[4]{\frac{4}{27}}\,\mu^{\frac34} \approx 0.620\,\mu^{\frac34}\ ,
\end{align}
where the inner and outer horizons coincide, and the black hole becomes extremal (but not supersymmetric). 
The associated Penrose diagram has only one asymptotic boundary which is repeated infinitely many times, see Figure~\ref{fig:ExtremalPenrose}.

When the charge is increased further so that $q \geq q_{\rm ext}$ the horizons disappear and one is left with a naked singularity.
Such solutions are usually discarded as non-physical based on the cosmic censorship conjecture \cite{Penrose:1969pc}.

\subsubsection*{Imaginary charge}

Up to now, we have discussed the case where the electric charge is real, $q\in \mathbb{R}$.
For later discussion, it is useful to examine the case of an imaginary charge. Consider $q = - i \tilde q$, with $\tilde q \in \mathbb{R}$, the main effect of which is to flip the sign in the last term of the blackening factor \eqref{eq:Blackening} and make the gauge field imaginary.\footnote{Such analytic continuations in charge were often discussed in the context of Euclidean solutions, see for example \cite{Hawking:1995ap}. However, solutions with imaginary charge are often discarded since the energy-momentum tensor is negative.} 
As a result, there is only one real solution to the equation~\eqref{eq:HorizonEq}, meaning that the resulting geometry has only one (outer) horizon, while the location of the inner horizon becomes imaginary (see~\eqref{rminus}), $r_- = - i \tilde r_-$.\footnote{The choice of the sign in the rotation of $q$ does not matter.}

\begin{figure}[t]
    \centering
    \includegraphics[width=0.5\linewidth]{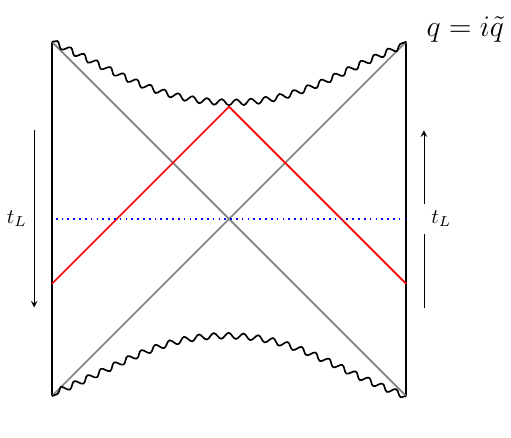}
    \caption{Penrose diagram for a black hole with imaginary charge. 
    With $t_L$ we denote the Lorentzian time and mark the  direction in which it increases on both boundaries.
    }
    \label{fig:PDImQ}
\end{figure}

The Penrose diagram now resembles that of a neutral black hole, see Figure~\ref{fig:PDImQ}, with only one horizon and a spacelike curvature singularity that is bend in compared to the boundaries.
In particular, the geometry contains an ``ordinary'' bouncing null geodesic, which can be straightforwardly connected to the boundary correlation functions, in exact parallel with the neutral case. Accordingly, we {\it define} the bouncing null geodesic in the charged black brane with real charge through analytic continuation from the imaginary-charge case, relating it to boundary correlation functions that are assumed to be analytic in $q$.

\subsection{Bouncing geodesics as a contour integrals}

In this subsection we study radial null geodesics in the charged black brane background \eqref{eq:ChargedMet1}. Before doing that, we note that Euclidean correlation functions of scalar operators in the large operator dimension limit can be obtained  by geodesics in the Euclidean section of the black brane geometry~\cite{Louko:2000tp, Fidkowski:2003nf, Festuccia:2005pi}. 
We discuss such geodesics in detail in Appendix~\ref{app:SpaceGeo} and their matching with the boundary OPE expansions. Here we focus on radial null geodesics in the Lorentzian geometry.

Consider a radial geodesic in the charged black brane background~\eqref{eq:ChargedMet1}. Due to the time isometry of the metric, we can introduce a conserved charge along a geodesic, 
\begin{align}
   {E} = r^2\, f(r)\,\dot t \,,
\end{align}
 which we refer to as the energy of the geodesic. The equation for a spacelike geodesic is given by 
 \be\label{space}
 -\frac{E^2}{r^2\,f(r)} + \frac{\dot r^2}{r^2\,f(r)} =1 \,,
\ee
while that for a null geodesic is
\begin{align} \label{null}
    \dot r^2 = E^2\,.
\end{align}
From~\eqref{null}, a radial null geodesic with $E>0$ that starts at the boundary reaches $r=0$, as in the lower half of the red line in Fig.~\ref{fig:PDRealQ}. Now suppose that such a null geodesic bounces off from the timelike singularity in a manner as indicated by the red line in the figure. The time shift between the endpoints of this null geodesic is given by
\begin{align}
\label{eq:TimeShiftDef}
\tau_c(\mu,q) = 2 i \int_0^{\infty} \frac{dr}{r^2 f(r)} , 
\end{align}
where we have included an additional factor of 
$i$ so that $\tau_c$ represents the {\it Euclidean} time shift, facilitating comparison with the CFT discussion later.

When $d$ is even, one can use the symmetry of the integrand
\begin{align}
    (-r)^2 f(-r) = r^2 f(r)\,,
\end{align}
to rewrite the shift as
\begin{align}
\label{eq:TimeShiftDef2}
    \tau_c(\mu,q) = i\int_{-\infty}^{\infty}\frac{dr}{r^2 f(r)}\,,
\end{align}
which allows us to  treat $r$ as a complex variable and evaluate this integral using contour integration.%
\footnote{For odd $d$, \eqref{eq:TimeShiftDef} can also be transformed into a contour integral, only that the integral from $-\infty$ to $0$ is taken along a ray at an angle of $2\pi/d$ with the pole along this line included in the wedge.}
Since the integrand has poles along the contour of integration, the final result depends on the chosen contour.
We will now present the results for a particular contour, which follows from analytic continuation from the imaginary $q$ case.

\begin{figure}[t]
    \centering
    \includegraphics[width=\linewidth]{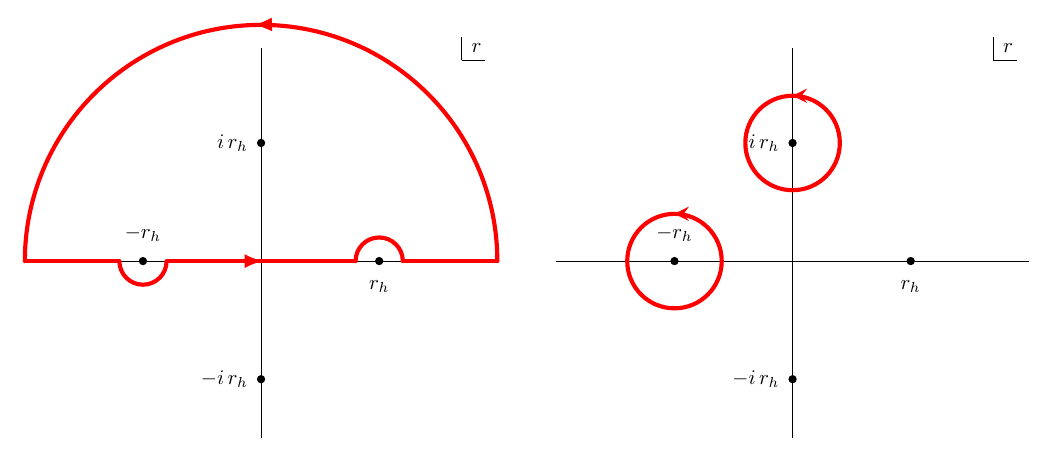}
    \caption{The contour which reproduces the time-shift of a bouncing geodesic reflecting off the future black hole  singularity in the case of a Schwarzschild-AdS black hole, where $r_h^4 = \mu$. Choosing the opposite prescription (picking up the poles at $+ir_h$ and $+r_h$) corresponds to the time-shift associated with the bounce off the past singularity.}
    \label{fig:4dUnchargedContour}
\end{figure}

It is instructive to start with the case of a neutral black hole, where $q=0$.
In this case, the choice of contour to obtain the  appropriate time shift in the case of spacelike geodesics was already discussed in \cite{Festuccia:2005pi}.
Extending these results to the case of null geodesics yields the contour depicted in Figure~\ref{fig:4dUnchargedContour}.
In the neutral case, we can write the time-shift as~\cite{Fidkowski:2003nf, Festuccia:2005pi, Ceplak:2024bja}
\begin{align}\label{e.defiieno}
    \tau_c(\mu,q=0) &= i \int_{\mathcal C}\,\frac{r^2\,dr}{r^4 - r_h^4} = - 2\pi \left[ \mathrm{Res}\left(\frac{r^2}{r^4 - r_h^4}, -r_h\right)+\mathrm{Res}\left(\frac{r^2}{r^4 - r_h^4}, ir_h\right)\right]\nonumber\\
    &= \frac{\pi}{\sqrt{2}\sqrt[4]{\mu}}e^{\frac{i\pi}{4}}\,,
\end{align}
where $r_h = \sqrt[4]{\mu}$, and $\mathcal{C}$ is the contour depicted in Figure~\ref{fig:4dUnchargedContour}.
The final result is indeed the result found in \cite{Fidkowski:2003nf, Festuccia:2005pi}. 

\begin{figure}[t]
    \centering
    \includegraphics[width=\linewidth]{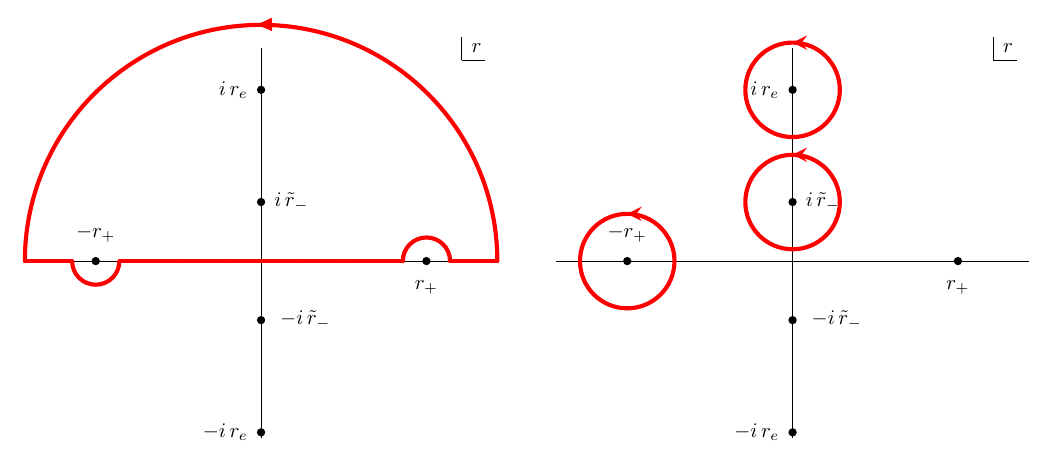}
    \caption{Contour chosen for the null geodesic in a charged black hole with imaginary charge ($q\in \mathrm{Im}$) in $d=4$. The contour picks up an additional contribution due to the extra pole in the upper-half plane due to the ``imaginary" inner horizon.}
    \label{fig:4dContStandard_Im}
\end{figure}

Next, consider the charged black hole with an imaginary charge. 
When $q\neq 0$, there are two additional poles in the complex-$r$ plane corresponding to the location of the inner horizon. 
If the charge is imaginary, these two new poles are on the imaginary axis. 
On the other hand, $r_+$ and $r_e$ smoothly change with the charge, as seen in \eqref{eq:RootExpansions}.
The contour in this case is the same as for the neutral case -- depicted in Figure~\ref{fig:4dContStandard_Im}.
But as there is an additional pole in the upper-half plane, the contour picks up an extra pole, the contribution of the imaginary inner horizon
\begin{align}
\label{eq:TimeShiftImQ}
    \tau_c(\mu, q=-i\qt) &= -2\pi\left[ \mathrm{Res}\left(\frac{1}{r^2f(r)}, -r_+\right)+\mathrm{Res}\left(\frac{1}{r^2f(r)}, +i \tilde r_-\right)+\mathrm{Res}\left(\frac{1}{r^2f(r)}, ir_e\right)\right]\nonumber\\
    &=\frac{\pi\,r_+^3}{(r_+^2 + \tilde r_-^2)(r_+^2 + r_e^2)}+ i\pi \left[\frac{\tilde r_-^3}{(\tilde r_-^2 + r_+^2)(\tilde r_-^2 - r_e^2)}+ \frac{r_e^3}{(r_e^2 + r_+^2)(r_e^2 - \tilde r_-^2)}\right] \nonumber\\*
    &= \frac{\tilde \beta_+}{2} + i\pi \left[\frac{\tilde r_-^3}{(\tilde r_-^2 + r_+^2)(\tilde r_-^2 - r_e^2)}+ \frac{r_e^3}{(r_e^2 + r_+^2)( r_e^2 - \tilde r_-^2)}\right] \,,
\end{align}
where $\tilde \beta_+$ is the temperature of the outer horizon given in \eqref{eq:Temp} and $r_- = -i \tilde r_-$. 
The real part of this expression is equal to half of the inverse temperature, consistent with the interpretation that such a null geodesic connects two points on opposite sides of the eternal black hole, as depicted in Figure~\ref{fig:PDImQ}.

Compared with the neutral case, the additional pole contributes only to the imaginary part of the Euclidean time shift, which gives the change in Lorentzian time between boundary points that are connected by the null geodesic.
Therefore, the contribution of the imaginary charge is to change the bend of the singularity compared to the boundary. 
Whether the singularity gets further bent in or less bent out depends on the sign of $\tilde q$. 

\begin{figure}[t]
    \centering
    \includegraphics[width=\linewidth]{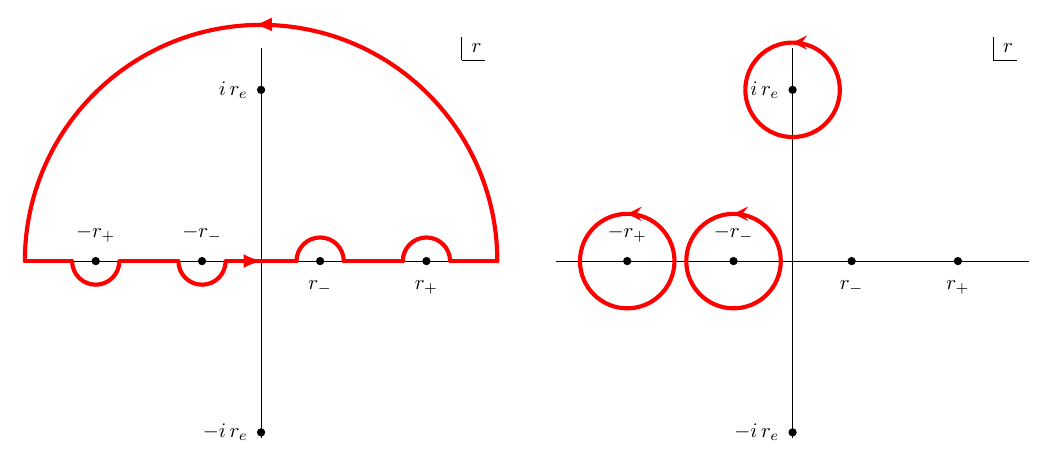}
    \caption{The choice of contour for calculating the time shift of bouncing geodesics in a charged black hole background with real charge in $d=4$.
    The choice of contour is consistent with the imaginary charge case, depicted in Figure~\ref{fig:4dContStandard_Im}, followed by analytically continuing $i \tilde r_- \to - r_-$. When closing the contour on the upper half plane, we pick up the residues at three poles: At $-r_+$, $-r_-$, and $ir_e$, which are then evaluated in Equation \eqref{eq:CriticalTauQ}.}
    \label{fig:4dContStandard}
\end{figure}

Now let us turn to the case of a real charge $q$, which can be obtained by rotating $\tilde r_- \to i r_-$, so that two simple poles appear along the real axis. In particular, the pole at $i \tilde r_-$ in the upper half-plane moves to the negative real axis. Analytic continuation from the imaginary-charge case then dictates that we choose the contour shown in Fig.~\ref{fig:4dContStandard}.

The time shift can then be found from the residues at the encircled poles
\begin{align}
\label{eq:CriticalTauQ}
    \tau_c(\mu,q) &= -2\pi\left[ \mathrm{Res}\left(\frac{1}{r^2f(r)}, -r_+\right)+\mathrm{Res}\left(\frac{1}{r^2f(r)}, -r_-\right)+\mathrm{Res}\left(\frac{1}{r^2f(r)}, ir_e\right)\right]\nonumber\\*
    &= \frac{\pi\,r_+^3}{(r_+^2\!- r_-^2)(r_+^2\!+ r_e^2)}+  \frac{\pi\,r_-^3}{(r_-^2\!- r_+^2)(r_-^2\!+  r_e^2)}+ i  \frac{\pi\,r_e^3}{(r_+^2+ r_e^2)(r_-^2+ r_e^2)}\nonumber\\*
    &= \frac{\beta_+ - \beta_-}{2} +  i  \frac{\pi\,r_e^3}{(r_+^2+ r_e^2)(r_-^2+ r_e^2)}\,,
\end{align}
 where we have expressed the results in terms of the inner and outer inverse temperature defined in \eqref{eq:Temp}.
The main observation is that the real part of $\tau_c(\mu,q)$, which gives the Euclidean time shift between the points connected by the bouncing geodesics, is not just $\beta_+/2$.
This means that the bouncing geodesics described by the contour in Figure~\ref{fig:4dContStandard} do not end up on the opposite side of the Penrose diagram, but on a different copy of the same side of the Penrose diagram, as depicted in Figure~\ref{fig:PDRealQ}.
The additional part, $- \beta_-/2$, can be attributed from the contributions from crossing the inner horizons.

While $\tau_c$ can be expressed in a closed form using $r_{\pm}$ and $r_e$, it is difficult to find an explicit closed form expression for this time shift in terms of $\mu$ and $q$. 
We can find it explicitly as a perturbative expansion in $q$, using~\eqref{eq:RootExpansions}, which gives 
\begin{align}
\label{eq:TauCExp1}
    \tau_c(\mu,q) &= \frac{ \pi }{\sqrt{2}
   \mu^{\frac14}}e^{\frac{i \pi }{4}}
   +\frac{3
   \pi \, q^2}{4 \sqrt{2}\,
   \mu ^{7/4}}e^{-\frac{i \pi }{4}}
   +\frac{45\pi \,
   q^4}{32 \sqrt{2} \,\mu ^{13/4}} e^{\frac{i \pi }{4}} 
   +\frac{385  \pi\,  q^6}{128 \sqrt{2} \,\mu^{19/4}}e^{-\frac{i \pi}{4}}\nonumber\\
   &\quad -\frac{\pi 
   \left(q^2\right)^{3/2}}{\mu ^{5/2}}
   +\frac{9
   \pi  \left(q^2\right)^{7/2}}{2 \mu ^{11/2}}
   +\coo{q^8}\,,
\end{align}
where the first line contains analytic contributions coming from the $r_+$ and $r_e$ residues in \eqref{eq:CriticalTauQ}, while the second line contains terms that are non-analytic in $q^2$ which stem only from the residues at the location of the inner horizon. 

It is also interesting to note the phase dependence of each term. 
The coefficients of all terms that are analytic in $q^2$ are complex, with a phase $e^{\pm\frac{i\pi}{4}}$, while the coefficients of non-analytic terms are all real. 
This has an interesting consequence when the expansion \eqref{eq:TauCExp1} is rewritten using the time shift of a bouncing geodesic in a neutral black hole~\eqref{e.defiieno}, after which
\begin{align}
\label{eq:TauCQExp2}
    \tau_c(\mu,q) & =  \tau_c(\mu,0) \le(1 + \frac{6\,\hat q^2}{\pi^6}\,+ \frac{32\,i\,\left(\hat q^2 \right)^{\frac32}}{\pi^9}\,- \frac{90\,\hat q^4}{\pi^{12}}\, + \ldots \ri) , \\
    \hat q^2 & \equiv q^2 \tau_c(\mu,0)^6 \ .
\end{align}
Here one notices that all $\mu$ and the complex phase dependence is captured by the appropriate power of $\tau_c(\mu,0)$ and the effective expansion parameter is the dimensionless number $\hat q^2 = q^2 \tau_c(\mu,0)^6$.
The remaining coefficients are just complex numbers: Interestingly, all prefactors of terms analytic in $q^2$ are real, while all terms that are non-analytic in $q^2$ are purely imaginary. 
This fact will become important later.

We obtained~\eqref{eq:CriticalTauQ} by analytically continuing the time shift of the bouncing null geodesic from the imaginary-charge case to the real-charge case, and interpreted it as the time shift of the bouncing null geodesic depicted in Fig.~\ref{fig:PDRealQ}.
A sceptical reader might wonder how we know that, after reaching the singularity, the null geodesic in Fig.~\ref{fig:PDRealQ} indeed bounces in the manner shown there.
The short answer is by symmetry under reflection in $t$.
In Fig.~\ref{fig:PDImQ}, the left half of the bouncing geodesic is obtained from the right half by reflecting in $t$, which, inside the horizon, interchanges left and right since $t$ is Euclidean there.
Upon analytic continuation to the real-charge case, the time coordinate $t$ in region~VII of Fig.~\ref{fig:PDRealQ} becomes Lorentzian, and reflection in $t$ now relates up and down.
Thus, the lower half of the null geodesic is reflected to obtain the upper half.

We can also obtain the bouncing null geodesic of Fig.~\ref{fig:PDRealQ} by considering the large-$E$ limit of a radial spacelike geodesic~\eqref{space}.
A radial spacelike geodesic with sufficiently large $E$ again reaches the timelike singularity and does not bounce.
Instead, we should consider the analytic continuation to real charge of the spacelike geodesic in the imaginary-charge case, whose large-$E$ limit gives the bouncing null geodesic of Fig.~\ref{fig:PDImQ}.
This analytic continuation yields a complex spacelike geodesic with a complex turning point near the singularity at $r=0$.
It can be shown that such a complex spacelike geodesic indeed bounces near the timelike singularity in the desired manner, and its $E \to \infty$ limit gives the null bouncing geodesic of Fig.~\ref{fig:PDRealQ}.
Further details are provided in Appendix~\ref{app:bounce}.

\subsection{Extremal black hole}

\begin{figure}[t]
    \centering
    \includegraphics[scale=0.6
]{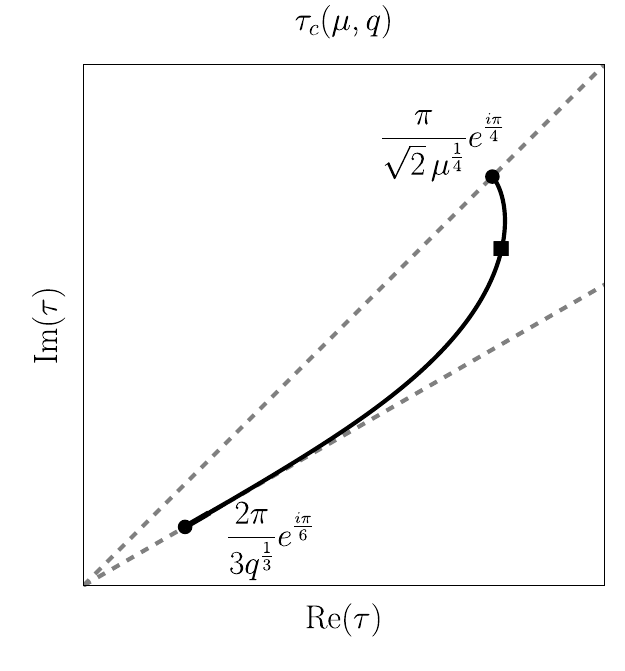}
    \caption{The behaviour of $\tau_c(\mu,q)$, given in \eqref{eq:CriticalTauQ}, in the upper-right quadrant of the complex-$\tau$ plane as a function of $q$ while keeping $\mu$ fixed. At $q=0$, the location of the singularity is at $\tau_c(\mu,0) = \frac{\pi}{\sqrt2\,\mu^{\frac14}}e^{\frac{i\pi}{4}}$ while at large $q$ the location of the pole is at $\tau_c \approx \frac{2\pi}{3\,q^{\frac13}
    }e^{\frac{i\pi}{6}}$. The two gray dashed lines represent fixed phases at $\tfrac{\pi}{2}$ and $\tfrac{\pi}{6}$. The time shift is finite and smooth across the extremal limit, denoted by black square, and for all values of the charge, $q$, regardless of whether the solution has a naked singularity or not.}
    \label{fig:TaucQNum}
\end{figure}

While the expansion \eqref{eq:TauCExp1} gives valuable insight into the analytic structure of the time shift as a function of the charge, it does not capture the full behaviour, especially near the extremal regime of the black hole. 
The expression \eqref{eq:CriticalTauQ} can be analysed numerically by changing the charge for fixed $\mu$, as depicted in Figure~\ref{fig:TaucQNum}.
The most prominent feature of this plot is that the time shift is a finite and smooth function of the charge. 
In particular, the extremal limit is smooth: As $q \to q_{\rm ext}$, $\tau_c$ stays finite.  
This can already be seen in \eqref{eq:CriticalTauQ}: The extremal limit is obtained when $r_- \to r_+$, at which point both $\beta_+$ and $\beta_-$ diverge. However,  their singular behaviour is exactly cancelled out in \eqref{eq:CriticalTauQ}, leaving only a finite result.%
\footnote{This can already be seen, for example, by considering $r_- = r_+ - \epsilon$ and find that as $\epsilon \to 0$, \mbox{$\tau_c = \pi\,r_+(r_+^2+ 3r_e^2)/(2(r_+^2 + r_e^2)^2)$}.}

It is important to note that one could also calculate $\tau_c(\mu, q_{\rm ext})$ by first taking $r_- \to r_+$ and then choosing the contour as in Figure~\ref{fig:4dContStandard}. 
In this case one picks up only two residues, one at $r = + i r_e$, where there is a simple pole, and one at $r = -r_+$, where one finds a double pole. However, the time shift calculated in this way is identical to the one obtained from \eqref{eq:CriticalTauQ}, where the extremal limit is taken only after the contour was chosen.%
\footnote{The statement of the commutativity of taking the extremal limit and choosing the contour can be seen from the fact that when the $r_-$ and $r_+$ poles collide, one does not pinch-off the contour, which leads to additional singular behaviour. }

This result suggests that bouncing geodesics can also probe the singularity of an extremal black hole. 
Since the real component of time shift is non-vanishing (see for example Figure~\ref{fig:TaucQNum}), this suggests that the bouncing geodesic still crosses a horizon, as shown in Figure~\ref{fig:ExtremalPenrose}. 
In this case, the bouncing geodesic starts at one copy of the boundary, crosses the degenerate horizon, reflects off the singularity, and after crossing the second degenerate horizon it ends on a second copy of the asymptotic boundary.

\subsection{Naked singularity}

When $q > q_{\rm ext}$, there are no horizons and thus the geometry contains a naked singularity. 
While such solutions are usually discarded on grounds of the loss of predictability of the theory, we can still analyse the behaviour of the time shift $\tau_c(\mu,q)$ in the super-extremal regime. 
One can already see from the numerical result in Figure~\ref{fig:TaucQNum}, that even in this regime of the parameter space, the time shift remains finite. 
Not only that, the real part is also non-vanishing, suggesting that the null geodesic crosses some kind of a horizon. 

\begin{figure}[t]
    \centering
    \includegraphics[width=\linewidth]{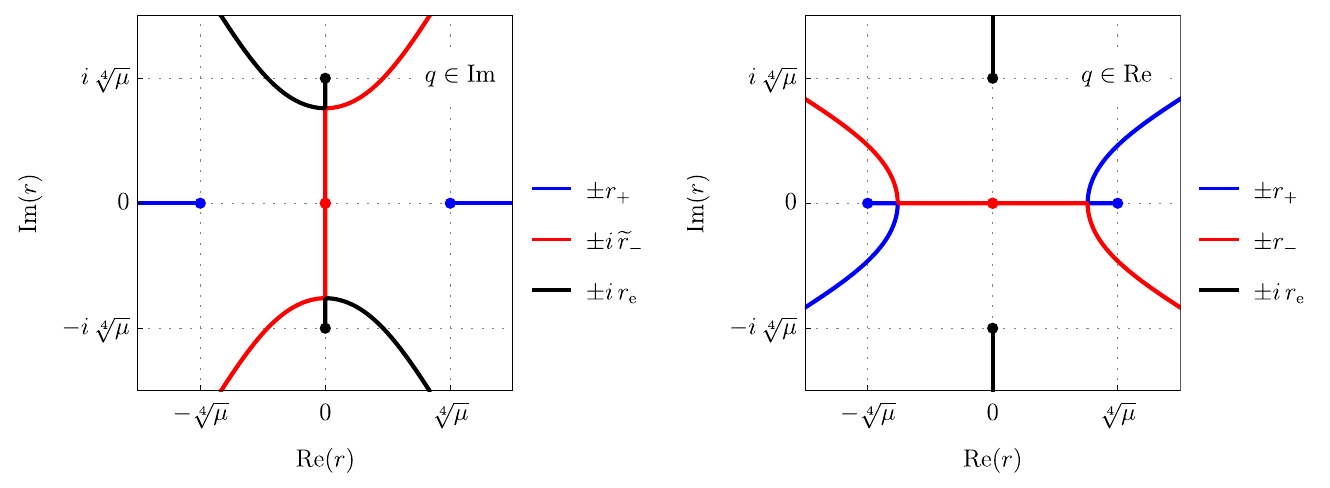}
    \caption{The location of the roots of the equation $f(r) = 0$ in the complex-$r$ plane as $q$ is changed while $\mu$ is kept fixed. 
    On the left, $q$ is imaginary, while on the right $q$ is real.
    On both plots, the dots correspond to the locations of the roots at $q=0$.
    In blue, we denote the movement of the outer horizon, $r_+$, while red depicts the movement of the root related to the inner horizon. The third root is denoted in black. 
    When the charge is real, the poles corresponding to the inner and outer horizon collide on the real line and then become complex. 
    This is not the case if the charge is imaginary, where the (outer) horizon is always on the real line, while $\tilde r_-$ and $r_e$ collide on the imaginary axis. Correspondingly, there is no extremal limit for the black hole with imaginary charge.}
    \label{fig:roots1}
\end{figure}

\begin{figure}[t]
    \centering
    \begin{subfigure}[b]{0.45\textwidth}
    \centering
    \includegraphics[width=\textwidth]{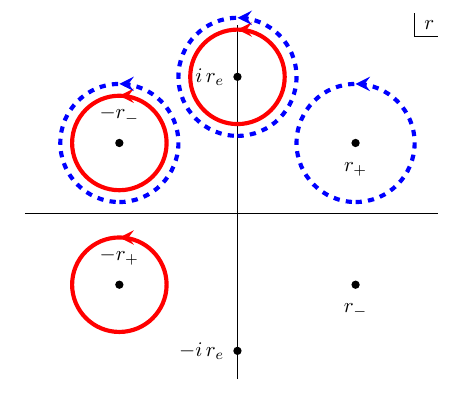}
  \end{subfigure}
  \begin{subfigure}[b]{0.45\textwidth}
    \centering
    \includegraphics[scale=0.9]{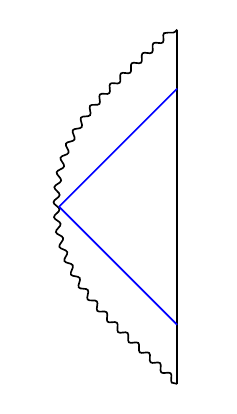}
  \end{subfigure}
    \caption{Left: The locations of the poles of the integral \eqref{eq:TimeShiftDef2} in the complex-$r$ plane for real charge in the regime $q> q_{\rm ext}$.
    In dashed blue we depict the collapsed contour obtained by integrating along the real line and closing the contour on the upper-half plane. The resulting time shift diverges at the extremal limit.
    The red circles enclose the poles that contribute to the time-shift which is calculated at $q < q_{\rm ext}$ and analytically continued past the extremal point. Such a time shift is smooth across the extremal point and is depicted in Figure~\ref{fig:TaucQNum}. Right: The time-shift $\tilde \tau_c$ associated with the dashed-blue contour is purely imaginary, suggesting that it can be interpreted as the time-shift of a null geodesic in a spacetime with a naked singularity in AdS.}
    \label{fig:Contour-naked}
\end{figure}

To see why this is the case, it is convenient to consider how the poles in the complex-$r$ plane move as the charge is increased while the mass of the black hole is kept fixed, see Figure~\ref{fig:roots1}.
Let us first consider the case of real charge.
When $q = 0$, there are four roots of the equation, two real and two imaginary. If we then turn on a small amount of charge two additional roots appear on the real axis, which correspond to the locations of the inner horizon. 
Meanwhile, the location of the outer horizon decreases while the poles on the imaginary axis move away from the origin. 
As the charge is increased, the locations of the inner and outer horizons grow closer until they collide at $q = q_{\rm ext}$ and move off the real axis so that for $q > q_{\rm ext}$, there is no real solution to $f(r) = 0$.
This means that schematically, the poles in the complex-$r$ for real charge $q > q_{\rm ext}$ are located as depicted in Figure~\ref{fig:Contour-naked}.

Continuation from $q < q_{\rm ext}$ dictates taking the contour shown on the right of Figure~\ref{fig:4dContStandard}, which computes the time shift as in \eqref{eq:CriticalTauQ}, and which is in turn smooth across the extremal point.
This is equivalent to picking up the poles encircled in red in Figure~\ref{fig:Contour-naked}, which  can be interpreted as a complex null geodesic connecting two copies of the naked singularity spacetime, and probes the black hole singularity.
The geodesic has to be complex, since its time-shift has a non-vanishing real component indicating crossing some horizon and there are no horizons on the real section of the spacetime for $q> q_{\rm ext}$.

Note that for an imaginary charge, there is no extremal limit. 
There exists a real bouncing geodesic for all values of the (imaginary) charge, including the regime with  $|\qt|> q_{\rm ext}$. 
At the ``extremal'' point,  the two poles along the imaginary axis in the complex-$r$ plane collide, see left of Figure~\ref{fig:roots1}.
The contour in the complex-$r$  for calculating the time-shift $\tau_c(\mu, -i\tilde q)$ therefore does not need to be modified and picks up the same residues as in \eqref{eq:TimeShiftImQ}.
One can alternatively obtain the contour for the real charge in the regime $q > q_{\rm ext}$ by analytic continuation from the imaginary-charge case, which again leads to the contour corresponding to picking up the residues at the poles encircled in red in Figure~\ref{fig:Contour-naked}.

We can calculate the behaviour of $\tau_c(\mu,q)$ in the large charge limit, $q^4 \gg \mu^3$, where the poles in the complex-$r$ plane are located at
\begin{subequations}
\label{eq:NakedPoles}
    \begin{align}
          r_+ &=
        e^{\frac{i \pi }{6}} q^{\frac13}-\frac{i \mu }{6 q}+\frac{e^{-\frac{ i
   \pi }{6}}\, \mu ^2}{72\, q^{7/3}}
   +\coo{q^{-3}}\,,\\
        r_- &=e^{-\frac{i\pi}{6}} q^{\frac13}+\frac{i \mu }{6
   q}+\frac{e^{\frac{i \pi }{6}} \,\mu ^2}{72\, q^{7/3}}+\coo{q^{-3}}\,,\\
       i  r_e &=i \left(q^{\frac13}+\frac{\mu }{6 q}-\frac{\mu ^2}{72 q^{7/3}}+\coo{q^{-3}}\right)\,,
    \end{align}
\end{subequations}
which gives 
\begin{align}
    \tau_c(\mu,q)\xrightarrow[q^4\gg \mu^3]{} \frac{2\pi}{3\,q^{\frac13}}e^{\frac{i\pi}{6}}- \frac{2\pi \mu}{18\,q^{\frac53}}\,e^{-\frac{i\pi}{6}}- \frac{i\pi\,\mu^2}{24\,q^3}+ \ldots\,.
\end{align}
One notes that this expression is perfectly analytic in $\mu$, which reflects the fact that in the super-extremal regime, varying $\mu$ does not significantly alter the geometry, in stark contrast to the addition of a small amount of charge which can drastically alter the causal structure of spacetime. 
Introducing
\begin{align}
\label{eq:qPoleDef}
    \tau_c(0,q) \equiv \frac{2\pi}{3\,q^{\frac13}}e^{\frac{i\pi}{6}}\,,
\end{align}
which is the leading pole at $\mu =0$ and we sometimes refer to as the ``$q$-pole", we can express 
the large-$q$ expansion as an expansion in dimensionless parameter  $\hat \mu \equiv \mu \tau_c^4 (0,q)$ as
\begin{align}
\label{eq:TauCSmallMu}
     \tau_c(\mu,q)\xrightarrow[q^4\gg \mu^3]{} \tau_c(0,q) \le(1 + \frac{27}{32\,\pi^4} \hat \mu + \frac{6561}{4096\,\pi^8}\,\hat \mu^2
   +\ldots \ri)\,.
\end{align}
Note that in this expansion all numerical prefactors are real.

To conclude this discussion, note that if one were to evaluate the integral \eqref{eq:TimeShiftDef2} along the real axis and close the contour in the upper half-plane,\footnote{The contribution from the semicircle at infinity again vanishes.} the poles in the upper half-plane---encircled by the dashed blue contour in Figure~\ref{fig:Contour-naked}---would be picked up, leading to
\begin{align}
\label{eq:TauSuperext}
    \tilde \tau_c(\mu,q) &= -\frac{\pi\,r_+^3}{(r_+^2\!- r_-^2)(r_+^2\!+ r_e^2)}+  \frac{\pi\,r_-^3}{(r_-^2\!- r_+^2)(r_-^2\!+  r_e^2)}+ i  \frac{\pi\,r_e^3}{(r_+^2+ r_e^2)(r_-^2+ r_e^2)}\, .
\end{align}
This expression is  purely imaginary for $q > q_{\rm ext}$ as $r_+^* = r_-$,\footnote{In this regime, labelling poles as $r_+$ and $r_-$ is arbitrary.} which means there is only real Lorentzian time shift and is consistent with a real null geodesic in the naked singularity spacetime in AdS, as depicted on the right of Figure~\ref{fig:Contour-naked}. The time shift corresponds to the separation in Lorentzian time between insertion points connected by such a bouncing geodesic.
Note that~\eqref{eq:TauSuperext} diverges in the extremal limit,  $r_- \to r_+$, because the sign of the first term has changed compared with~\eqref{eq:CriticalTauQ}. 
%

%%%%%%%%%%%%%%%%%%%%%%%%%%%%%%%%%%%%%%%%%%%%%%%%%%%%%
%%%%%%%%%%%%%%%%%%%%%%%%%%%%%%%%%%%%%%%%%%%%%%%%%%%%%

\section{Singularity in the OPE}
\label{sec:chargedbh}

In this section we analyze thermal two-point functions of scalar operators in a state at finite temperature and charge. 
We decompose the correlators using the OPE and compute the coefficients associated with the exchange of stress-tensor and current composites.
We determine the asymptotic behavior of these OPE coefficients, which allows us to resum the expansion and show that the stress-tensor and charged sectors exhibit  a singularity in the complex $\tau$ plane independent of the conformal dimension $\Delta <\infty$. 
We provide support that this singularity can be  related to the null geodesics that bounces off the time-like singularity of the charged black hole in AdS.

%%%%%%%%%%%%%%%%%%%%%%%%%%%%%%%%%%%%%%%%%%%%%%

\subsection{Holographic calculation and operator structure}
\label{ssec:qneq0ansatz}

Our aim is to compute the two-point function
    \begin{equation}\label{Cor}
   G(\tau)=\expval{\phi(\tau)\phi(0)}_\beta\ ,
    \end{equation}
for a {\it neutral} scalar operator $\phi$ of dimension $\Delta$ in the boundary theory at finite temperature and charge density, which on the gravity side is described by a charged black brane.

For $q=0$, the boundary correlator $G(\tau)$ can be decomposed into the stress-tensor sector and the double-trace sector contributions, as discussed below~\eqref{decom}. 
In the charged case an additional conserved current $J_\mu$ with spin $1$ and conformal dimension 3 is involved. 
Primary operators can then be constructed from products of these currents, $e.g.$ $J_{\mu} J_{\nu} \ldots J_{\rho}$, together with all possible index contractions.\footnote{For a neutral scalar operator $\phi$, only operators formed from even number of $J_\mu$ 
can appear in the OPEs of $\phi \phi$ due to conservation of $U(1)$ charge.}
We will encounter mixing with the stress-tensor sector; for example, $[T^3]_{J'=4}$ has the same scaling dimension and spin as the operator $J_{\mu} J_{\nu} J_{\rho} J_{\lambda}$. One can also form primary operators by combining stress tensors and currents; see 
Tab.\ \ref{tab:t2}.
\begin{table}[t!]
    \centering
    \begin{minipage}{0.45\textwidth}
        \centering
        \begin{tabular}{|c|c|c|}
            \hline
            \,\,$\Delta'$\,\, & Operator & \,\,Spin\,\, \\ \hline
            0 & $\mathds{1}$ & 0 \\ \hline
            4 & $T_{\mu\nu}$ & 2 \\ \hline
            6 & $\mqty{{\color{reed}J_\mu J^\mu} \\ {\color{reed}J_\mu J_\nu}}$ & $\mqty{{\color{reed}0}\\{\color{reed}2}}$ \\ \hline
            8 & $\mqty{T_{\mu\nu}T^{\mu\nu} \\ T_{\mu\nu}T^{\nu}_{\phantom{\nu}\rho} \\ T_{\mu\nu}T_{\rho\lambda}}$ & $\mqty{0\\2\\4}$ \\ \hline
            10 & $\mqty{{\color{reed}T_{\mu\nu}J^\mu J^\nu}\\{\color{reed}T_{\mu\nu}J^{\nu}J_{\rho}}\\{\color{reed}T_{\mu\nu}J_\rho J_\lambda}}$ & $\mqty{{\color{reed}0}\\{\color{reed}2}\\{\color{reed}4}}$ \\ \hline
        \end{tabular}
    \end{minipage}%
    \hspace{-0.12\textwidth}
    \begin{minipage}{0.45\textwidth}
        \centering
        \begin{tabular}{|c|c|c|}
            \hline
            \,12\, & $\mqty{T_{\mu\nu}T^\nu_{\phantom{\nu}\rho}T^{\rho\mu}\phantom{\big|}\\{\color{reed}J^\mu J_\nu J_{\mu}J^\nu}\\T_{\mu\nu}T^{\nu}_{\phantom{\nu}\rho}T^{\rho}_{\phantom{\rho}\lambda}\\{\color{reed}J_\mu J_\nu J_\rho J^\rho}\\T_{\mu\nu}T_{\rho\lambda}T^{\lambda}_{\phantom{\lambda}\alpha}\\{\color{reed}J_\mu J_\nu J_\rho J_\lambda}\\T_{\mu\nu}T_{\rho\lambda}T_{\alpha\beta}}$ & $\mqty{\,\,\,\,0\,\,\,\,\\{\color{reed}0}\\2\\{\color{reed}2}\\4\\{\color{reed}4}\\6}$ \\ \hline
            14 & $\mqty{{\color{reed}T_{\mu\nu}T^{\nu}_{\phantom{\nu}\rho}J^{\rho}J^{\mu}}\\{\color{reed}T_{\mu\nu}T_{\rho\lambda}J^\nu J^\lambda}\\{\color{reed}T_{\mu\nu}T^{\nu}_{\phantom{\nu}\rho}J_\lambda J_\alpha}\\{\color{reed}T_{\mu\nu}T_{\rho\lambda}J_\alpha J_\beta}}$ & $\mqty{{\color{reed}0}\\{\color{reed}2}\\{\color{reed}4}\\{\color{reed}6}}$ \\ \hline
        \end{tabular}
    \end{minipage}
    \caption{Schematic form of the relevant primary operators up to dimension $\Delta'=14$. Operators novel to the charged case are highlighted in red.}
    \label{tab:t2}
\end{table}

It is then convenient to decompose~\eqref{Cor} into contributions from a single \textit{stress-charge sector}, $G_{\rm T+J}(\tau)$, which includes all operators generated by the stress tensor and the current, together with the double-trace sector.
Decomposing $G_{\rm T+J}$ in the OPE \eqref{e.tcbdecoomp} one finds
\begin{align}
\label{eq:OPEres1}
    G_{\rm T+J}(\tau) = \frac{1}{\tau^{2\Delta}}\left(1+\sum_{n=0}^{\infty}\Lambda_n(\mu, q)\,\tau^{4+2n}\right)\,.
\end{align}

To obtain the CFT data $\Lambda_n(\mu, q)$, we follow the approach introduced in~\cite{Fitzpatrick:2019zqz}, where they can be extracted from the solutions to the equation of motion for the bulk field dual to~$\phi$, via a suitable near-boundary expansion.
Although this method was originally constructed for holographic theories where stress-tensor is the only conserved charge, we can modify the ansatz for the case of a charged black brane -- see Appendix \ref{app:OPE} for details.  In $d=4$ the first few coefficients read:
\begin{subequations}
\label{e.Lb00st}
    \begin{align}
    \Lambda_0(\mu,q)&=\frac{\Delta}{40}\,\mu, \\
    \Lambda_1(\mu,q)&=\frac{9 \Delta+11 \Delta ^2-5 \Delta ^3}{840(\Delta-3)(\Delta-2)}\,q^2, \\
    \Lambda_2(\mu,q)&=\frac{144 \Delta-88 \Delta ^2+672 \Delta ^3-413 \Delta ^4+63 \Delta ^5}{201600 (\Delta -4) (\Delta -3) (\Delta -2)}\,\mu^2, \\
    \Lambda_3(\mu,q)&=-\frac{840 \Delta +1482 \Delta ^2+991 \Delta ^3+248 \Delta ^4-376 \Delta ^5+55 \Delta ^6}{369600 (\Delta -5) (\Delta -4) (\Delta -3) (\Delta -2)}\,\mu\,q^2 \ . 
    \end{align}
\end{subequations}

Since the bulk metric~\eqref{eq:ChargedMet1} depends only on $q^2$, and the neutral bulk field~$\phi$ does not couple to the $U(1)$ gauge field~$A$, the coefficients $\Lambda_n(\mu,q)$ are polynomial functions of $q^2$.
We will now proceed to analyse the OPE data $\Lambda_n(\mu,q)$ in detail.

%%%%%%%%%%%%%%%%%%%%%%%%%%%%%%%%%%%%%%%%%%%%%%

\subsection{Leading order in the small-\texorpdfstring{$q$}{q} expansion}
\label{ssec:MuPole}

We first examine the behavior of~\eqref{eq:OPEres1} at the leading order in small $q$ expansion, which already yields a rich amount of information.

Expand the stress-tensor and the charge correlator perturbatively in $q^2$
\begin{align}
\label{eq:CorrPertExp}
    G_{\rm T+J}(\tau) = \frac{1}{\tau^{2\Delta}}\Bigg[\sum_{n=0}^{\infty}\,\mu^{n}\,\Lambda_n^{(0)}(\Delta)\,\tau^{4n} + q^2 \sum_{n=0}^{\infty}\,\mu^{n}\,\Lambda_n^{(2)}(\Delta)\,\tau^{6+4n}+\coo{q^4}\Bigg]\,,
\end{align}
where we have taken out the appropriate power of $\mu$, and the resulting  coefficients $\Lambda_n^{(k)}(\Delta)$ are only functions of the conformal dimension $\Delta$. 
One finds the following asymptotic behaviour for the OPE coefficients 
{\small{
\begin{align}
    \label{eq:AsyAns0}
    \Lambda_n^{(0)}(\Delta) \xrightarrow[]{n\rightarrow\infty} c(\Delta) \, \frac{n^{2\Delta-3}}{\left[-\left(\frac{\pi}{\sqrt{2}}\right)^4\right]^n}\,, \qquad \Lambda_n^{(2)}(\Delta)\xrightarrow[]{n\rightarrow\infty} -4\,c(\Delta)\,c_2(\Delta)\,\frac{n^{2\Delta-2}}{\left[-\left(\frac{\pi}{\sqrt{2}}\right)^4\right]^n}\,.
\end{align}}}%
The asymptotic expression for $\Lambda_n^{(0)}$ was  found in \cite{Ceplak:2024bja}. 
The new result is the expression for $\Lambda_n^{(2)}$.
Here we note three important features: (1) The denominator remains the same as  $\Lambda_n^{(0)}$;
(2) $\Lambda_n^{(2)}$ carries an additional factor of $n$ compared to $\Lambda_n^{(0)}$;
(3) We isolated the same $c(\Delta)$ prefactor as in $\Lambda_n^{(0)}$, 
which was determined in \cite{Ceplak:2024bja}, 
\begin{align}
\label{eq:tildeCDef}
    c(\Delta) = \frac{\Delta}{40}\,\tilde c(\Delta) \times 2\,\,c_f(\Delta)\,4^{2\Delta}\,,\qquad \tilde c(\Delta) \equiv\frac{\pi\,\Delta\,(\Delta-1)}{\sin\left(\pi\,\Delta\right)\,\Gamma\left(2\Delta+\frac32\right)}\,,
\end{align}
with $c_f(\Delta)$ a slowly varying function of $\Delta$ that is approximately unity. $c_2 (\De)$ will be discussed below.

The contribution from the large $n$-part of the sum to~\eqref{eq:CorrPertExp} can be approximated by inserting~\eqref{eq:AsyAns0}
into the expansion and replacing the sum by an integral,
\begin{align}
\label{eq:AsySum}
     \frac{c(\Delta)}{\tau^{2\Delta}}
    \int_0^{\infty}\,dn\,n^{2\Delta-3}\,\left(\frac{\tau^4}{\tau_c(\mu,0)^4}\right)^{n}\,\bigg[1-4\,c_2(\Delta)\,q^2\,n\,\tau^6 +\coo{q^4}\bigg]\,,
\end{align}
where we introduced%
\footnote{It is important to note that $\tau_c(\mu,0)^4 = - \frac{\pi^4}{4\mu}\in \mathbb{Re}$. Since $c(\Delta)$ and $c_2(\Delta)$ are also real, the integral \eqref{eq:AsySum} is real when $\tau\in \mathbb{R}$. However, $\tau_c(\mu,0)$ is complex and only expanding $\tau$ near \eqref{eq:TauCq0} moves us into the complex plane.}
\begin{align}
    \label{eq:TauCq0}
    \tau_c(\mu,0) = \frac{\pi}{\sqrt{2}\,\mu^{\frac14}}e^{\frac{i\pi}{4} + \frac{i\,\pi\,k}{2}}\,,\qquad k =0,1,2,3\,.
\end{align}
The integrals in~\eqref{eq:AsySum} can be evaluated using
\begin{align}
    \label{eq:GenInt}
    I_m(y)\equiv \int_0^{\infty}\,y^{n}n^{2\Delta+m-1}\,dn = \Gamma(2\Delta+m)\left(-\log y\right)^{-(2\Delta+m)}\,,
\end{align}
from which~\eqref{eq:AsySum} becomes 
\begin{align}
\label{eq:CorrEval0}
     \frac{c(\Delta)}{\tau^{2\Delta}}\left[I_{-2}\left(\frac{\tau^4}{\tau_c(\mu,0)^4}\right)
    - 4 \,c_2(\Delta)\,q^2\,\tau^6\,I_{-1}\left(\frac{\tau^4}{\tau_c(\mu,0)^4}\right) 
  + \coo{q^4}
    \right]\,.
\end{align}

We will now use the above large-$n$ contribution to approximate $G_{\rm T+J}(\tau)$, which is valid 
around the  singular point closest to the origin,
$i.e.$, as $\tau \to \tau_c(\mu,0)$. 
Expanding $\tau$ around $\tau_c(\mu,0)$, we find 
\begin{align}
\label{eq:CorrNearMuSingu}
    G_{\rm T+J}(\tau) \approx 
    \frac{c(\Delta)\,\Gamma(2\Delta-2)}{4^{2\Delta-2}\,\tau_c(\mu,0)^2}\,\frac{1}{\left(\tau_c(\mu,0)-\tau\right)^{2\Delta-2}}\Bigg[1
    -c_2(\Delta)\,(2\Delta-2)\,q^2\,\frac{\tau_c(\mu,0)^7}{\tau_c(\mu,0)-\tau}
    + \ldots\Bigg]\,,
\end{align}
where only the leading terms in the small $(\tau_c(\mu,0)-\tau)$ expansion of the two contributions in~\eqref{eq:CorrNearMuSingu} have been kept.

Notice that the $q^2$ term contains an additional factor of $(\tau_c(\mu,0) - \tau)^{-1}$. The form~\eqref{eq:CorrNearMuSingu} has a natural interpretation in terms of a shifted $\tau_c$. More explicitly, consider 
\begin{align}
\label{eq:SinguExp2}
\frac{1}{\left[\left(\tau_c(\mu,0) +\de \tau_c \right) -\tau\right]^{2\Delta-2}}\approx \frac{1}{\left(\tau_c(\mu,0)-\tau\right)^{2\Delta-2}}\Bigg[1
    -(2\Delta-2)\frac{\de \tau_c}{\tau_c(\mu,0)-\tau}
    + \ldots\Bigg]\, .
\end{align}
We can match~\eqref{eq:CorrNearMuSingu} with~\eqref{eq:SinguExp2} in terms of a shifted $\tau_c (\mu, q)$ as 
\be \label{newtau}
\tau_c (\mu, q) = \tau_c(\mu,0) +\de \tau_c, \quad \de \tau_c = q^2 \,c_2(\Delta) \, \tau_c(\mu,0)^7 + \cdots \ .
\ee
In performing the expansion~\eqref{eq:SinguExp2}, we work in the regime $\delta\tau_c \ll \tau - \tau_c(\mu,0)$, which can be realized for sufficiently small $q$, such that $\tau - \tau_c(\mu,0)$ remains within the range where our approximation~\eqref{eq:CorrNearMuSingu} is valid.

\begin{figure}[t]
    \centering
    \includegraphics[scale=1]{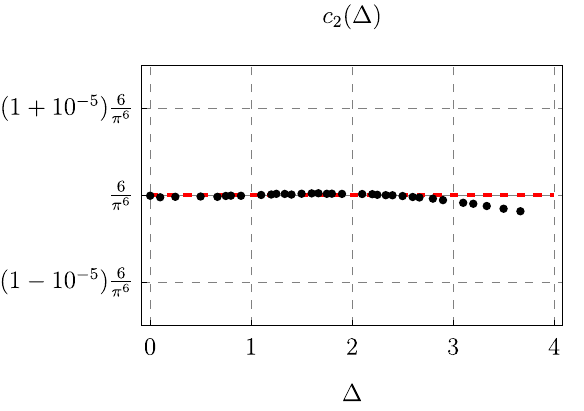}
    \caption{Comparison between $c_2(\Delta)$ obtained from the asymptotic analysis of $\Lambda_n^{(2)}(\Delta)$ (black markers) and the geodesic coefficient $6/\pi^6$ (red dashed). In the range $\Delta \in (0,4)$, the numerical value obtained from the analysis of the OPE coefficients and the geodesic result agree to at least five digits. 
    The details of how the $c_2(\Delta)$ was obtained is discussed in Section~\ref{ssec:NonAnalytic}.} 
    \label{fig:c2comp}
\end{figure}

Equation~\eqref{newtau} can now be compared with the $q^2$ term in the time-shift~\eqref{eq:TauCExp1}, obtained from the bouncing null geodesic in the charged black hole geometry. We find a match provided that $c_2(\Delta) =  6/\pi^6$. 
The values of $c_2(\Delta)$ can be readily calculated from the numerical data.
We plot them in Figure~\ref{fig:c2comp}. 
We find that with our data (where the OPE data in \eqref{eq:CorrPertExp} is truncated at $n_{\rm max} = 220$), for low-lying $\Delta$, where the numerical analysis is stable, $c_2(\Delta)$ agrees with the geodesic prediction with high precision. 
The drop-off in accuracy as $\Delta$ is increased is related to the cross-over behaviour discussed in \cite{Ceplak:2024bja}. We believe that if $n_{\rm max}$ is increased, the accuracy at higher $\Delta$ is increased as well.

That, at large $n$, $\Lambda_n^{(2)}(\Delta) \sim n \Lambda_n^{(0)}(\Delta)$ implies that the expansion of $\Lambda_n(\mu, q)$ in powers of $q^2$ takes the form
\begin{equation}
\Lambda_n(\mu, q) \sim \Lambda_n(\mu, 0) + \mathcal{O}(q^2 n) + \cdots ,
\end{equation}
which in turn means that, for any fixed (however small) $q$, the subleading correction eventually exceeds the leading term at sufficiently large $n$, and the expansion in $q$ breaks down.
This signals possible non-analytic behavior in $q^2$ around $q = 0$.
Understanding such non-analyticity requires studying the large-$n$ behavior of $\Lambda_n^{(m)}(\Delta)$ for general $m$.
We will return to this issue in Sec.~\ref{ssec:NonAnalytic}.

%%%%%%%%%%%%%%%%%%%%%%%%%%%%%%%%%%%%%%
%%%%%%%%%%%%%%%%%%%%%%%%%%%%%%%%%%%%%
\subsection{The small-\texorpdfstring{$\mu$}{mu} expansion}
\label{ssec:qpole}

Understanding the general behavior of $\Lambda_n(\mu, q)$ is numerically very challenging.
Before addressing it directly, as another warm-up, in this subsection we consider its behavior in the opposite regime: large $q$, with $\mu$ treated as a small perturbation.
In this regime, the bulk geometry~\eqref{eq:ChargedMet1} exhibits a naked singularity.
Nevertheless, from the boundary perspective, the OPE expansion can, in principle, be extended to this regime.
Since the procedure described in Appendix~\ref{app:OPE} concerns only the asymptotic region of the bulk geometry, our discussion applies straightforwardly here as well.

Expanding the OPE coefficients in small-$\mu$, we have 
\begin{align}
\label{eq:CorrMuExp}
    G_{\rm T+J}\approx \frac{1}{\tau^{2\Delta}}\Bigg[\sum_{n=1}Q_n^{(0)}(\Delta)\,q^{2n}\,\tau^{6n} + \mu\,\tau^4 \sum_{n=1}Q_n^{(1)}(\Delta)\,q^{2n}\,\tau^{6n}+\coo{\mu^2}\Bigg]\,.
\end{align}
The first term in the brackets corresponds to the $G_{\rm J}(\tau)$ sector, which contains exchanges involving only charge-current operators, while the second term, proportional to $\mu$, represents the exchange of a single stress-tensor operator together with an arbitrary number of charge-current operators.

We are again interested in the large-$n$ asymptotic behaviour of the coefficients $Q_n^{(i)}(\Delta)$, which can be found to be
\begin{subequations}
        \label{eq:QnAsy}
    \begin{align}
        Q_n^{(0)}(\Delta) &\sim c^{(q)}(\Delta)\,\left(-\frac{3^6}{(2\,\pi)^6}\right)^n\,n^{2\Delta-\frac{14}{5}}\,,\\
        Q_n^{(1)}(\Delta) &\sim- 6\, c^{(q)}(\Delta)\,c_2^{(q)}(\Delta)\,\left(-\frac{3^6}{(2\,\pi)^6}\right)^n\,n^{2\Delta-\frac{9}{5}}\,,
    \end{align}
\end{subequations}
where the numerical prefactors of the above expressions are parametrized for later convenience.
We again observe that the subleading term has one higher power in $n$ than the leading term. 

Inserting these asymptotic expressions into~\eqref{eq:CorrMuExp} and approximating the sum with an integral gives
\begin{align}
    G_{\rm T+J} \approx \frac{c^{(q)}(\Delta)}{\tau^{2\Delta}}\int_0^{\infty}\,dn\,n^{2\Delta-\frac{14}{5}}\,\left(\frac{\tau^6}{\tau_c(0,q)^6}\right)^n
    \left[1- 6\,c_2^{(q)}(\Delta)\,\mu\,n\,\tau^4+ \coo{\mu^2}\right]\,,
\end{align}
where we have defined 
\begin{align}
\label{eq:QPoleDef}
    \tau_c(0,q) \equiv \frac{2\pi}{3q^{\frac13}}\,e^{\frac{i\pi}{6} + \frac{i\,\pi\,k}{3}}\,, \qquad k = 0,1,\ldots 5\,.
\end{align}
Using the integral \eqref{eq:GenInt}, we can evaluate the expansion as 
\begin{align}\label{uenw}
   G_{\rm T+J} \approx \frac{c^{(q)}(\Delta)}{\tau^{2\Delta}}\Bigg[I_{-\frac{9}{5}}\left(\frac{\tau^6}{\tau_c(0,q)^6}\right)-  6\,c_2^{(q)}(\Delta)\,\mu\,\tau^4\,I_{-\frac{4}{5}}\left(\frac{\tau^6}{\tau_c(0,q)^6}\right)+ \coo{\mu^2}\Bigg]\,.
\end{align}
As in the previous subsection, the above expression should be trusted only near $\tau\to \tau_c(0,q)$, where it is singular.
Expanding~\eqref{uenw} to leading order in $(\tau_c(0,q)-\tau)$ at each order in the $\mu$-expansion, we  find
{\small{
\begin{align}
\label{eq:CorrNearQSingu1}
    G_{\rm T+J}(\tau) \approx \frac{c^{(q)}(\Delta)\,\Gamma\left(2\Delta-\frac95\right)}{6^{2\Delta-\frac95}\,\tau_c(0,q)^{\frac95}}\,\frac{1}{\left(\tau_c(0,q)\!-\!\tau\right)^{2\Delta-\frac95}}\Bigg[1
    -c^{(q)}_2(\Delta)\,\left(2\Delta-\frac95\right)\,\mu\,\frac{\tau_c(0,q)^5}{\tau_c(0,q)-\tau}
    + \ldots\Bigg]\ .
\end{align}}}%
It is interesting to note that the order of the leading singularity is $2\Delta - 9/5$, which is greater than $2\Delta-2$ of \eqref{eq:SinguExp2}.

\begin{figure}[t]
    \centering
    \includegraphics[scale= 1]{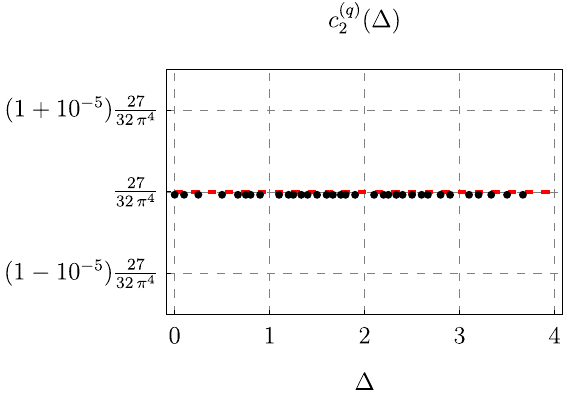}
    \caption{Comparison between the numerical values of $c_2^{(q)}$ (black markers) obtained from the asymptotic analysis of the OPE coefficients and the value predicted by the null geodesic $27/(32\pi^4)$ (red dashed). The two values agree for $\Delta \in (0,4)$ to at least five digits. }
    \label{fig:qpoleNumbers2}
\end{figure}

We can again rewrite~\eqref{eq:CorrNearQSingu1} in the form 
\be
G_{\rm T+J}(\tau)  \propto {1 \ov \left(\tau_c(\mu,q)\!-\!\tau\right)^{2\Delta-\frac95}} ,
\ee
with 
\be  \label{eq:QPoleExp}
\tau_c (\mu, q) = \tau_c (0, q) + \de \tau_c, \quad \de \tau_c = 
c^{(q)}_2(\Delta)\,\mu\, \tau_c(0,q)^5  \ .
\ee
Comparing with~\eqref{eq:TauCSmallMu}, we see that~\eqref{eq:QPoleExp} coincides with the time-shift of the bouncing null geodesic provided that  
\begin{align}
    c_2^{(q)} (\Delta) = \frac{27}{32\,\pi^4}\,.
\end{align}
Indeed, in Figure~\ref{fig:qpoleNumbers2}, we show that from our numerical analysis, the matching is quite accurate for low enough $\Delta$.

We also note that
\begin{align}
    c^{(q)}(\Delta) \approx \frac{\Delta}{40}\,\tilde c(\Delta) \, 2\,c_f^{(q)}(\Delta)\,6^{2\Delta}\,,
\end{align}
where $c_f^{(q)}(\Delta)$ is a function of order 1 with a mild $\Delta$-dependence and $\tilde c(\Delta)$ was defined in \eqref{eq:tildeCDef}.

\subsection{General \texorpdfstring{$\mu$}{mu} and \texorpdfstring{$q$}{q}}

We now turn to the large-$n$ asymptotic behavior of $\Lambda_n(\mu, q)$ for general $\mu$ and $q$, and discuss its implications.
Given the number of parameters involved, it is no longer possible to reliably deduce an analytic asymptotic form of $\Lambda_n(\mu, q)$ for large $n$ from numerical data.
An alternative strategy must therefore be employed.

In the preceding two subsections, we found agreement between the singularity location $\tau_c$ of $G_{\rm T+J}$ and the null-geodesic time shift in two opposite regimes: near $q = 0$ with finite $\mu$, and near $\mu = 0$ with finite $q$.
Encouraged by this success, it is natural to conjecture that, for general $\mu$ and $q$, $G_{\rm T+J}$ exhibits a singularity of the form
\begin{align} \label{conjM}
\frac{1}{\left(\tau_c(\mu,q)-\tau\right)^{2\Delta - b(\mu,q)}} ,
\end{align}
where $\tau_c(\mu,q)$ is given the time shift of the null geodesic (see Figure~\ref{fig:TaucQNum}), and $b(\mu,q)$ is some function with 
$b (\mu, 0) =2$ and $b(0, q) = {9 \ov 5}$.

\begin{figure}[t]
    \centering
    \includegraphics[scale =1
]{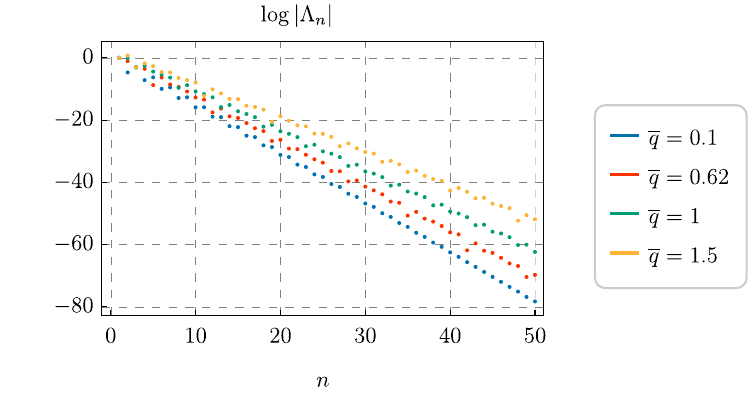}
    \caption{Numerical values for $\log|\Lambda_n(\overline q)|$ at different values of $\overline q$. In all cases, the conformal dimension of the insertion operator is taken to be $\Delta = 3/2$.
    It is important to note that $\overline q = 0.62$ is very close to the extremal point $\overline q_{\rm ext} = (4/27)^{\frac14}\approx 0.6204$. This shows that the OPE data close to extremality does not exhibit any diverging behaviour.}
    \label{fig:RawDataFixedq}
\end{figure}

We will now proceed to provide support for this conjecture. For this purpose, 
it is  convenient to rewrite the OPE expansion as
\begin{align}
\label{eq:GenForm}
G_{\rm T+J}(\tau) = \frac{1}{\tau^{2\Delta}}\left(1+ \frac{\Delta\,\mu\,\tau^4}{40}\sum_{n=0}^{\infty}\,\mu^{\frac{n}{2}}\Lambda_n(\overline q)\,\tau^{2n}\right)\,.
\end{align}
In the above expression, we have singled out the contribution of the identity and expressed the factor $\frac{\Delta\,\mu\,\tau^4}{40}$, which is associated with the exchange of the single stress-tensor sector and whose OPE coefficient is protected by symmetry.
The coefficients $\Lambda_n(\overline q)$ are functions of the dimensionless parameter\footnote{In this subsection we suppress the dependence of the OPE coefficients on the conformal dimension.}
\begin{align}
\label{eq:qtdef}
    \overline{q} = \frac{q}{\mu^{\frac34}} \ .
\end{align}
From the structure of the OPE expanion \eqref{eq:OPEres1}, $\Lambda_n(\overline q)$ are polynomials  in $\overline q$ of degree at most $n$.
The data for $\Lambda_n(\overline q)$ at four representative values of $\overline q$ and $\Delta = \tfrac{3}{2}$ are shown in Figure~\ref{fig:RawDataFixedq}.
We consider the following ansatz for the asymptotic large-$n$ behavior of $\Lambda_n(\overline q)$,
\begin{align}
\label{eq:LambdaqtAns1}
\Lambda_n(\overline q) \sim \tilde c(\Delta) \, f_{\Delta}(\overline q) \, \frac{\cos\Big(2n \varphi_c(\overline q) - \psi_{\Delta}(\overline q)\Big)}{|\tau_c(\overline q)|^{2n}}\,(2n)^{2\Delta - b_{\Delta}(\overline q) - 1}\ ,
\end{align}
which is motivated by the fact that, when substituted into~\eqref{eq:GenForm}, it leads to the form \eqref{conjM}, as we will show below.
$\tilde c(\Delta)$ was introduced earlier in~\eqref{eq:tildeCDef}, and the functions $f_{\Delta}(\overline q)$, $b_{\Delta}(\overline q)$, $\psi_{\Delta}(\overline q)$, $\varphi_c(\overline q)$, and $|\tau_c(\overline q)|$---to be determined---are all taken to be real.
Note that~\eqref{eq:LambdaqtAns1} is not a polynomial in $\overline q$; rather, it should be viewed as a continuum approximation to $\Lambda_n(\overline q)$ when $n$ is treated as a continuous parameter in the large-$n$ limit.

We now show that~\eqref{eq:LambdaqtAns1} leads to~\eqref{conjM}. 
The form \eqref{eq:LambdaqtAns1} is manifestly real, but can be rewritten as a sum of two complex exponential functions 
\begin{align}
    \label{eq:LambdanAns2}
     \Lambda_n(\overline q)
   &\sim  \frac12\,\tilde c(\Delta) f_{\Delta}(\overline q) (2n)^{2\Delta-b_{\Delta}(\overline q)-1}\Bigg[\frac{e^{i\psi_{\Delta}(\overline q)}}{\left(|\tau_c(\overline q)|\,e^{i\varphi_c(\overline q)}\right)^{2n}} + \frac{e^{-i\psi_{\Delta}(\overline q)}}{\left(|\tau_c(\overline q)|\,e^{-i\varphi_c(\overline q)}\right)^{2n}}\Bigg]\nonumber\\*
     &=  \frac12\,\tilde c(\Delta) f_{\Delta}(\overline q) (2n)^{2\Delta-b_{\Delta}(\overline q)-1}\Bigg[\frac{e^{i\psi_{\Delta}(\overline q)}}{\tau_c(\overline q)^{2n}} + \frac{e^{-i\psi_{\Delta}(\overline q)}}{\tau^*_c(\overline q)^{2n}}\Bigg]\,,
\end{align}
where going into the last line we introduced
\begin{align}
\label{eq:CritPointEuler}
    \tau_c(\overline q) \equiv|\tau_c(\overline q)|\,e^{i\varphi_c(\overline q)}\,.
\end{align}
Inserting~\eqref{eq:LambdanAns2} into~\eqref{eq:GenForm} and approximating the sum with an integral gives
\begin{align}
    G_{\rm T+J}(\tau) \approx \frac{1}{\tau^{2\Delta}}\Bigg\lbrace1\!+\! \frac{\Delta\mu\tau^4}{80}\tilde c(\Delta) f_{\Delta}(\overline q)\int_0^{\infty}\!\!\dd n \,(2n)^{2\Delta-b_{\Delta}(\overline q)-1}\bigg[\frac{e^{i\psi_{\Delta}(\overline q)}}{\tau_c(\mu,q)^{2n}} + \frac{e^{-i\psi_{\Delta}(\overline q)}}{\tau^*_c(\mu,q)^{2n}}\bigg]\tau^{2n}\Bigg\rbrace\,,
\end{align}
where we introduced 
\begin{align}
\label{eq:CritPoint2}
    \tau_c(\mu,q) \equiv \frac{\tau_c(\overline q)}{\mu^{\frac14}}\,.
\end{align}
Evaluating the integral gives
\begin{equation}
\begin{split}
    G_{\rm T+J}(\tau)& \approx \frac{1}{\tau^{2\Delta}}\Bigg\lbrace
    1
    + \frac{\Delta\mu\tau^4}{80}
    \tilde c(\Delta) f_{\Delta}(\overline q) \Gamma(2\Delta-b_{\Delta}(\overline q))\,\,\times\\
    &
    \hspace{-30pt}\left[e^{i\psi_{\Delta}(\overline q)}\left(\!-\log\left(\frac{\tau^2}{\tau_c(\mu,q)^2}\right)^{-(2\Delta-b_{\Delta}(\overline q))}\right) + 
    e^{-i\psi_{\Delta}(\overline q)}\left(\!-\log\left(\frac{\tau^2}{\tau^*_c(\mu,q)^2}\right)^{-(2\Delta-b_{\Delta}(\overline q))}\right) 
    \right]
    \Bigg\rbrace.
\end{split}
\end{equation}
This expression is singular at four values of $\tau$: $\pm \tau_c(\mu, q)$ and $\pm \tau^*_c(\mu, q)$.
Expanding around $\tau_c(\mu,q)$ for concreteness, we find that 
\begin{equation}
\label{eq:GeneralSingu}
    G_{\rm T+J}(\tau\to \tau_c(\mu,q)) \approx \frac{\Delta\,\mu}{40}\,\frac{\tilde c(\Delta) f_{\Delta}(\overline q) \Gamma(2\Delta-b_{\Delta}(\overline q))\,e^{i\psi_{\Delta}(\overline q)}}{2^{2\Delta-b_{\Delta}(\overline q)+1}}\,\frac{\tau_c(\mu,q)^{4-b_{\Delta}(\overline q)}}{\left(
    \tau_c(\mu,q)-\tau\right)^{2\Delta-b_{\Delta}(\overline q)}}\,.
\end{equation}
Analogous expressions can be found for the other singular points of \eqref{eq:LambdanAns2}.
For generic values of $\mu$ and $q$, the ``residues'' at these singular points will be complex, and will depend on the value of the functions $(\psi_{\Delta}(\overline q),  b_{\Delta}(\overline q),f_{\Delta}(\overline q) )$.
This is in contrast to the two cases that we have already explored in the previous subsections, see \eqref{eq:CorrNearMuSingu} and \eqref{eq:CorrNearQSingu1}, where the residues at the singular points are real.

Ideally, one would take the ansatz~\eqref{eq:LambdaqtAns1}, fit it to the numerical data for $\Lambda_n(\overline q)$ obtained from the holographic calculation (such as those shown in Fig.~\ref{fig:RawDataFixedq}), extract the values of all the functions appearing in the ansatz, and demonstrate that the time shift of the null geodesic is reproduced from $\tau_c(\mu, q)$.
Unfortunately, we were not able to carry out this procedure in a reliable manner, due to the large number of functions involved. 
Instead, we fix $\tau_c(\mu, q)$ by hand using the time shift of the null geodesic and fit the remaining functions.
As we will show below, this approach yields a reliable fit.

\begin{figure}[t]
    \centering
    \includegraphics[width=\linewidth]{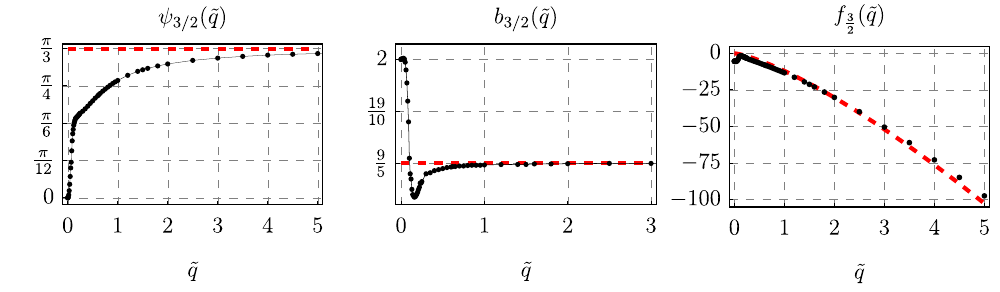}
        \caption{
    The behaviour of the functions $\psi_{3/2}(\overline q)$ (left), $b_{3/2}(\overline q)$ (middle), and $f_{3/2}(\overline q)$ (right) as a function of the dimensionless parameter $\overline q ={q}/{\mu^{\frac34}}$.
    The phase $\psi_{3/2}(\overline q)$ starts at the origin and goes to $\pi/3$ for large values of the charge.
    The exponent $b_{3/2}(\overline q)$ starts at $2$, which is the value observed for the uncharged black hole and tends to $9/5$, the value we find for the $q$-pole.
    Finally, the prefactor $f_{3/2}(\overline q)$ grows as $q^{\frac43}$ at large $\overline q$. %
     The red dashed line is the behaviour predicted by the $q$-pole analysis which predicts that for large charge, this prefactor function should behave as $f_{\Delta}(\overline q) \approx - 12.\, \overline q^{\frac43}$.
     For comparison, the best fit for the numerical data is at large $\overline q$  approximately $f_{\Delta}(\overline q) \approx - 11.4\, \overline q^{\frac43}$.
    }

    \label{fig:numfits1}
\end{figure}

\begin{figure}[t]
    \centering
    \includegraphics[width=\linewidth]{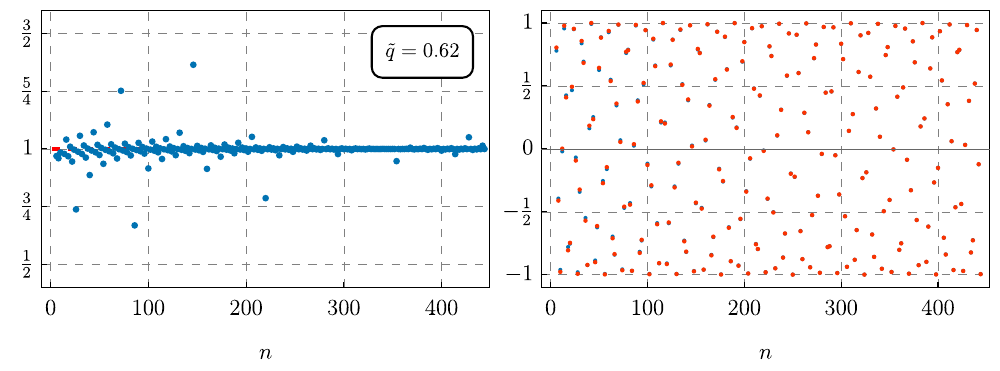}
    \caption{Left: The ratio between the numerical data $\Lambda_n(\overline q)$ and the asymptotic model given in \eqref{eq:LambdaqtAns1}. Right: From the numerical data $\Lambda_n(\overline q)$ we divide out the phase independent part of the asymptotic ansatz. We can then compare the actual data (blue markers) with the asymptotic prediction for the phase (red markers). We note that for large $n$, the asymptotic model incredibly accurately predicts all points. In both these plots $\overline q = 0.62$, which is extremely close to extremality. This shows that the asymptotic model given in \eqref{eq:LambdaqtAns1} works even in the extremal limit. Similar plots can be generated for all values of $\overline q$.
    }
    \label{fig:NumEvidence}
\end{figure}

The fit  for $\Delta = \frac32$ is presented in Figure~\ref{fig:numfits1}, and in Figure~\ref{fig:NumEvidence}, 
we show that the fit leads to excellent and highly non-trivial agreement between the ansatz and the actual numerical values. 
The functions $\psi_{\Delta}(\overline q)$, $b_{\Delta}(\overline q)$, and $f_{\Delta}(\overline q)$ all seem to be reasonably smooth functions of $\overline q$. 
While we cannot explain the non-trivial features that appear in the region of $\overline q \in \left[0,1\right)$, we can check whether the values at $\overline q = 0$ and $\overline q \gg1$ agree with the analysis in the preceeding subsections.

For $\Delta= 3/2$, we find that (see Figure~\ref{fig:numfits1})
\begin{subequations}
    \begin{align}
    &\overline q =0: && \psi_{\frac32}(0) = 0\,, \qquad  b_{\frac32}(0)= 2\,, \qquad f_{\frac32}(0) \approx -5.3\,,\\
    &\overline q \gg 1 : && \psi_{\frac32}(\overline q) \to \frac{\pi}3\,, \qquad  b_{\frac32}(
    \overline q
    )\to  \frac{9}{5}\,, \qquad f_{\frac32}(\overline q) \approx - 11.4 \,\overline q^{\frac43}\,.
\end{align}
\end{subequations}
The first line corresponds to the leading order term found in \eqref{eq:CorrNearMuSingu}, with the vanishing phase shift and the $b_{\frac32}(0) = 2$ are consistent with \eqref{eq:CorrNearMuSingu}.
To match the value of $f_{\frac32}(0)$, we need to compare the prefactors in \eqref{eq:CorrNearMuSingu} and \eqref{eq:GeneralSingu}. 
Once all the common factors are cancelled out, the former predicts that
\begin{align}
 f_{\frac32} (0) = -\frac{2^9}{\pi^4}c_f\left(\frac32\right) \approx - 5.26\,,
\end{align}
which is consistent with what is observed from numerical fits. 

The $\overline q \gg 1$ regime needs to match the pole we found in \eqref{eq:CorrNearQSingu1}.
Indeed, we immediately recognise $b_{\frac32}= 9/5$ as the shift in the exponent of the singularity.
To match the remaining two factors, we must compare the prefactors of \eqref{eq:CorrNearQSingu1} and \eqref{eq:GeneralSingu}.
After cancelling out all the common factors, one is left with
\begin{align}
    f_{\frac32}(\overline q) \,e^{i\psi_{\frac32}(\overline q)}  =  \frac{2^{5-9/5}c_f^{(q)}(3/2)\,6^{9/5}}{\mu\,\tau_c(0,q)^4} =- \frac{2^{5-9/5}c_f^{(q)}(3/2)\,6^{9/5}}{\left(\frac{2\pi}{3}\right)^4}\,\overline q^{\frac43}\,e^{\frac{i\,\pi}{3}}\approx -12.0 \,\overline q^{\frac43}\,e^{\frac{i\,\pi}{3}}\,.
\end{align}
One finds that the phase is matched to high accuracy, however, the best-fit value is off by about 5\%.

\subsection{Non-analytic behaviour in \texorpdfstring{$q$}{q} and inner horizons}
\label{ssec:NonAnalytic}

We have provided evidence that the location $\tau_c(\mu, q)$ of the singularity~\eqref{conjM} in $G_{\rm T+J}(\tau)$ can be identified with the time shift of the bouncing null geodesic.
In this subsection, we assume this identification to be correct and explore the analytic structure of the OPE coefficients $\Lambda_n(\mu, q)$ in~\eqref{eq:OPEres1}.
The bouncing geodesic not only probes the timelike singularity but also crosses both the outer and inner horizons.
In particular, we will identify the analytic structure associated with the presence of the inner horizon.

The small-$q$ expansion of $\tau_c(\mu, q)$ given in~\eqref{eq:TauCExp1} and~\eqref{eq:TauCQExp2} takes the form of a double series expansion,
\begin{align} \label{tauX}
\tau_c (\mu, q) &= \tau_0 + q^2 \tau_1 + q^4 \tau_2 + \cdots \cr
&\quad + (q^2)^{3 \ov 2} \le( \tau_{3 \ov 2} + q^2 \tau_{5 \ov 2} + \cdots \ri) ,
\end{align}
where the subseries proportional to $(q^2)^{3 \ov 2}$ arises solely from the contribution of the inner horizon.

We now examine the origin of the two subseries in the asymptotic large-$n$ behavior of $\Lambda_n(\mu, q)$.
For this purpose, it is convenient to expand $\Lambda_n(\mu, q)$ in powers of $q^2$, as in~\eqref{eq:CorrPertExp}, schematically,
\be\label{ehn1}
\Lambda_n(\mu, q) \sim \Lambda_n^{(0)} + q^2 \Lambda_n^{(2)} + \cdots + q^{2k} \Lambda_n^{(2k)} + \cdots ,
\ee
and to analyze the large-$n$ behavior of the coefficients $\Lambda_n^{(2k)}$.

In Sec.~\ref{ssec:MuPole}, we discussed that the $\tau_1$ term in~\eqref{tauX} can be seen from that 
$\Lam_n^{(2)}/\Lam_n^{(0)} \propto n$. More explicitly, the presence of $\tau_1$ term leads to the following subsequence in the $q^2$ expansion of~\eqref{conjM} (with $b(\mu,0) =2$), 
\be \label{ehn2}
{1 \ov (\tau_0 + q^2 \tau_1 - \tau)^{2\De-2}} = \frac{1}{\left(\tau_0 -\tau\right)^{2\Delta - 2}} \le(1 + \# {\tau_1 q^2 \ov \tau_0 - \tau} + \cdots 
+ \# \le({\tau_1 q^2 \ov \tau_0 - \tau} \ri)^k + \cdots \ri) \ .
\ee
The $q^{2k}$ term in~\eqref{ehn2} implies that, in~\eqref{ehn1}, $\Lambda_n^{(2k)}$ must contain a contribution such that $\Lambda_n^{(2k)} / \Lambda_n^{(0)} \propto n^k$, with an appropriate numerical prefactor to reproduce the corresponding term in~\eqref{ehn2}. We have checked that $\Lam_n^{(4)}$ is indeed proportional to $n^2$ at large $n$, with a coefficient precisely consistent with the above expansion.\footnote{Note that for an algebraic singularity of the form~\eqref{ehn2}, the order of doing the $n$ and $k$ sums does not matter. We have performed the sum over $n$ first and then the $k$-sum. One could also first sum over the $q$-expansion in~\eqref{eq:CorrPertExp} first for each $n$ and then sum over $n$.}  

Similarly, the $\tau_m$-term in~\eqref{tauX} for an integer $m$ implies that
$\Lambda_n^{(2km)}/ \Lambda_n^{(0)} $ should contain a contribution proportional to $n^k$ for $k =1,2,\cdots$. 
In other words, the higher-order terms in the first line of~\eqref{tauX} require specific subleading contributions in $n$ within $\Lambda_n^{(2k)}$.
For instance, in the ratio $\Lambda_n^{(4)} / \Lambda_n^{(0)}$, in addition to the leading term of order $n^2$ (corresponding to $\tau_1$), there should exist a specific subleading term of order $n$~(corresponding to $\tau_2$).

The fractional series in~\eqref{tauX} is more subtle. The $\tau_{3 \ov 2}$ term in~\eqref{tauX} naively leads to the following subsequence in the small-$q$ expansion of~\eqref{conjM}, 
\be \label{ehn3}
\frac{1}{\left(\tau_0 -\tau\right)^{2\Delta - b(\mu,0)}} \le(1 + \# {\tau_{3 \ov 2} (q^2)^{3 \ov 2} \ov \tau_0 - \tau} + \cdots 
+ \# \le({\tau_{3 \ov 2} (q^2)^{3 \ov 2} \ov \tau_0 - \tau} \ri)^k + \cdots \ri) ,
\ee
 inconsistent with the OPE expansion, which involves only even powers of $q$.
Here, we propose a possible scenario for how such a contribution may arise.

In the discussion of the integer subseries in~\eqref{tauX}, we have already seen that the subleading terms (in $n$) in $\Lambda_n^{(2k)}$ play an important role. To understand the origin of terms in~\eqref{ehn3}, we should thus understand more systematically the subleading terms, which we will start with $\Lam_n^{(0)}$ and $\Lam_n^{(2)}$. 

In \eqref{eq:AsyAns0}, we only present the leading order behavior in $n$. In principle there are other, subleading contributions, and in \cite{Ceplak:2024bja} we assumed that these come in an $1/n$ expansion.
This was motivated by the series expansion in the two-dimensional case where the thermal correlator is know exactly and these subleading corrections can be tracked.
In higher dimensions this seems to not be the case.
Indeed, one can observe%
\footnote{This was first observed by \cite{Afkhami-Jeddi:2025wra}. We would like to thank Simon Caron-Huot, Robin Karlsson, and Matthew Dodelson for discussions on this point.}
that the convergence properties of the asymptotic coefficients improves when one considers
\begin{align}
\label{eq:AsyAns1}
    \Lambda_n^{(0)}(\Delta) \sim c(\Delta) \, \frac{n^{2\Delta-3}}{\left[-\left(\frac{\pi}{\sqrt{2}}\right)^4\right]^n}
    \left[1+ \sum_{k=1}^{\infty}\frac{p_k(\Delta)}{n^k} +\sum_{l=1}^{\infty}\frac{r_l(\Delta)}{n^{\frac43\,l}} \right]\,,
\end{align}
where in addition to a $1/n$ expansion, one also considers a $1/n^{4/3}$ expansion. 
The addition of these powers is motivated by the analysis of the first non-trivial correction to the asymptotic behaviour of quasi-normal modes $\omega_n$ at large $n$ \cite{Musiri:2005ev, Dodelson:2023vrw}, a connection which we will discuss in detail elsewhere.
Repeating the same analysis at order $q^2$, one finds that the subleading terms of $\Lam_n^{(2)}$ can be written as
\begin{align}
    \label{eq:AsyAns2}
    \Lambda_n^{(2)}(\Delta)\sim -4\,c(\Delta)\,c_2(\Delta)\,\frac{n^{2\Delta-2}}{\left[-\left(\frac{\pi}{\sqrt{2}}\right)^4\right]^n}\left[1+ 
   {s_{1 \ov 3} (\De) \ov n^{1 \ov 3}} + \cdots \ri]
\end{align}
And indeed, one finds that including these subleading terms improves the numerical accuracy of the fitted values.
For example, Figure~\ref{fig:c2comp} has been computed using truncated versions of the expansions \eqref{eq:AsyAns1} and \eqref{eq:AsyAns2}.

From~\eqref{eq:AsyAns1} and~\eqref{eq:AsyAns2}, we find that 
\be 
q^2 {\Lambda_n^{(2)} \ov \Lambda_n^{(0)}} = q^2 \le(n + n^{2 \ov 3} + \cdots \ri)  \ .
\ee
We discussed before that the $\mathcal{O}(n)$ term leads to $\tau_1$ shift in~\eqref{tauX}. Now we show that $n^{2 \ov 3}$ term could lead to the $\tau_{3 \ov 2}$ term in~\eqref{tauX}. Combining the $\Lam_n^{(0)}$ term and the $n^{3 \ov 2}$ part of the $\Lam^{(2)}$ term, and approximating their contributions by integrals as before, we find  
\begin{align}
   &  
    \int_0^{\infty}\,dn\,n^{2\Delta-3}\,\left(\frac{\tau^4}{\tau_0^4}\right)^{n}\,\bigg[1-4\,c_2(\Delta)\, s_{1 \ov 3} (\De) \,q^2\,n^{2 \ov 3} \,\tau^6 \bigg] \\
    & =   I_{-2}\left(\frac{\tau^4}{\tau_0^4}\right) -4\,c_2(\Delta)\, s_{1 \ov 3} (\De) \,q^2\,\tau^6 I_{-\frac43}\left(\frac{\tau^4}{\tau_0^4}\right) \\
    & = {C \ov (\tau_0 - \tau)^{2 \De-2}} \le(1 - a {q^2 \ov (\tau_0 - \tau)^{2 \ov 3}} + \cdots \ri),
    \label{newX1}
    \end{align}
where, in the last line, we have kept only the leading contributions obtained by expanding each term around $\tau_0$, and
\be 
C= \Ga (2 \De-2), \quad a = 4^{1 \ov 3} {\Ga (2 \De-{4 \ov 3}) \ov \Ga (2 \De -2)} \,c_2(\Delta)\, s_{1 \ov 3} (\De) \tau^{6+{2 \ov 3}}_0 \ .
\ee

We imagine that when the higher order terms in $q^2$ are included, equation~\eqref{newX1} can be summed into a function 
$ f (\tau)$ which  has an algebraic singularity of the form 
\be
{1\ov (\tau_0 + (q^2)^{3 \ov 2} \tau_{3 \ov 2} - \tau)^{2 \De-2}}  \ .
\ee
One such example is 
\be
{C \ov \le((\tau_0 - \tau)^{2 \ov 3} + {a q^2 \ov 3 (\De -1)} \ri)^{3 (\De-1)}} + \cdots ,
\ee
but there could be many others. At the moment, our numerical data appear not enough to pinpoint the form of the function $f(\tau)$.\footnote{Unfortunately, this simplest example does not appear to fit well with our numerics.}

The presence of a non-trivial exponent function $b (\mu, q)$ in~\eqref{conjM} implies further subleading behavior (in $n$) in $\Lam_n^{(2k)}$. Assuming 
that $b (\mu, q)$ has a regular expansion in $q^2$ as $b(\mu, q)= b_0 + q^2 b_1 + \cdots$ where $b_0 = 2$. Then its expansion leads to a term of the form 
\be 
{1 \ov (\tau_0 - \tau)^{2\De-2}} \le(1 + b_1 q^2 \log (\tau_0 - \tau) + \cdots \ri)  \ .
\ee
The logarithmic term above can be generated by the presence of a $\log n$ term in $\Lam^{(2)}_n$ (and higher orders from higher powers of $\log n$ in $\Lam^{(2k)}_n$). Other possible dependence on $q$ in the prefactor of~\eqref{conjM} (not written explicitly there) would require specific $\mathcal{O}(1)$ terms in $\Lam^{(2)}_n$ and higher order coefficients. Checking these statements is numerically much more challenging and will be left for the future.

\section{Discussion and outlook}
\label{sec:Disc} 
%% Disucssion

We have analysed two-point correlation functions of scalar fields in the charged black brane background. 
In particular, we determined the asymptotic form of the OPE coefficients associated with the exchange of stress-tensor and current operators, as well as their composites. 
By resumming this part of the OPE, we showed that this sector develops divergences at specific locations in the complex time plane that coincide with the time shift of a bouncing geodesic in the charged black hole geometry: a null geodesic that starts at one asymptotic boundary, reflects off the timelike singularity, and returns to another asymptotic region on the same side of the Penrose diagram, see Figure~\ref{fig:PDRealQ}. 

It is important to emphasise that the OPE data associated with stress-tensor and current exchanges depends only on the near-boundary region of the Euclidean section of the geometry. 
The analytic continuation of the T+J sector of the Euclidean correlation function to the complex time plane enables us to probe many features of the bulk spacetime beyond the outer horizon, including signatures of the inner horizon and the singularities.

Our results challenge the conventional wisdom that the regions inside the inner horizons cannot be captured by the boundary theory.
In particular, the boundary system appears to ``know'' about the time-like singularities and the additional asymptotic boundaries hidden behind the inner horizons.
At the level of our discussion, the bulk scalar field dual to the boundary scalar operator acts purely as a probe and does not backreact on the geometry.
On the boundary side, we are working with the two-point function in a special state---the thermofield double state at finite temperature and finite chemical potential---in the large $N$ and large $\lam$ limit, whose analytic continuation is expected to fully match that of the bulk.
Thus, as already mentioned in the Introduction, our findings are {\it not} in tension with the expectation that the inner horizons are unstable and evolve into singularities under physical perturbations, as such effects should appear at order $1/N$.
It is of clear importance to explore boundary signatures of inner-horizon instabilities,\footnote{See~\cite{Shahbazi-Moghaddam:2024emr,Chen:2024ojv} and references therein for some recent discussions.} and our results provide a starting point for such investigations.

Surprisingly, we find that the presence of the inner horizon can be probed by terms that are non-analytic in $q^2$.
We also provide an argument that this non-analyticity in $q^2$ is, in turn, connected to the subleading behavior of the large-order OPE coefficients.
It would be instructive to understand the physical origin of these observations more clearly.
One may also ask whether the inner horizon is associated with an analogous non-analytic behavior in the angular momentum of rotating black holes~\cite{Hawking:1998kw,Levi:2003cx,Gibbons:2004js}.

We have shown that the location of the singularity in the corresponding OPE sector varies smoothly across the extremal point. 
For the boundary theory on the sphere, near the extremal limit, large quantum fluctuations of certain metric components are expected, which can significantly modify the classical geometry~\cite{Iliesiu:2020qvm}, including the region near the singularity \cite{Lin:2022rzw}.%
\footnote{See for example \cite{Emparan:2025sao, Emparan:2025qqf, Biggs:2025nzs} and references therein for observable signatures of such strong quantum effects.}
Our results are {\it not} in tension with these findings, as we are working in the probe limit at the leading order in the large $N$ expansion (and with the boundary theory on the plane, where such effects are absent). 
It would be interesting to understand  the effects of such large fluctuations by coupling the scalar operators to the Schwarzian subsector~\cite{Maldacena:2016upp, Engelsoy:2016xyb, Jensen:2016pah, Mertens:2017mtv, Kitaev:2018wpr, Yang:2018gdb, Iliesiu:2019xuh}.

We find that the boundary correlation function can probe even the regime where the bulk geometry has a naked singularity via complex null geodesics. It would be valuable to explore this direction further, especially in holographic scenarios where naked singularities can form dynamically \cite{Emparan:2021ewh}. 
A natural question for future work is which CFT observables could serve as a litmus test for determining whether the dual bulk geometry contains a naked singularity.

In this paper, we have studied the simplest observable---two-point functions of neutral scalar fields. 
More refined observables may reveal a richer structure. 
For example, charged scalar two-point functions have been shown in the semiclassical approximation to display a more intricate behaviour \cite{Brecher:2004gn}.%
\footnote{The semiclassical analysis of \cite{Brecher:2004gn} revealed a rich structure of correlation functions as the charge of the black hole increases, including a non-analytic behaviour at a finite value of the charge. It would be interesting to understand how the features discussed there emerge from the finite-$\Delta$ analysis presented in our work.}

The boundary OPE approach could potentially offer a pathway toward a holographic description singularities in more ``realistic'' geometries, such as the BKL chaotic dynamics \cite{Mis69, BKL70}. 
Recent progress in this direction includes \cite{Frenkel:2020ysx, Hartnoll:2020fhc, Hartnoll:2020rwq, DeCHat23, Carballo:2024hem, Caceres:2024edr, Arean:2024pzo}.%
\footnote{An interesting observation was made in \cite{Oling:2024vmq}, where the ultralocal behaviour of spacetime near curvature singularities was shown to be captured by the Carroll limit of general relativity.}
Finally, the singularities identified in the T+J sector of the correlation function may provide a means to study stringy and quantum resolutions of curvature singularities through field-theoretic methods. 
In particular, one may consider lower-dimensional $N \times N$ matrix models which, in the large-$N$ limit, possess a gravitational bulk dual. 
Correlation functions at finite $N$ would then serve as a toy model for quantum resolutions of curvature singularities, even in the absence of a well-defined bulk interpretation \cite{CERNgroup}. 
%

%%%%%%%%%%%%%%%%%%%%%%%%%%%%%%%%%%%%%%%%%%%%%%%%%%%%%
%%%%%%%%%%%%%%%%%%%%%%%%%%%%%%%%%%%%%%%%%%%%%%%%%%%%%

\section*{Acknowledgements}

We would like to thank Ilija Burić, Simon Caron-Huot, Alejandra Castro, Matthew Dodelson, Rober\-to Emparan, Chantelle Esper, Blaise Goutéraux, Ivan Gusev, Michal P. Heller, Cristoforo Iossa, Robin Karlsson,  Manuela Kulaxizi, Adam Levine, Gerben Oling  for discussions.
The work of N\v{C} and AP is supported in part by Taighde \'Eireann - Research Ireland under the grant agreement 22/EPSRC/3832. SV was supported in part by the
Irish Research Council Government of Ireland Postgraduate Fellowship under project
award number GOIPG/2023/3661. HL is supported by the Office of High Energy Physics of U.S. Department of Energy under grant Contract Number  DE-SC0012567 and DE-SC0020360 (MIT contract \# 578218), and  by grant \#63670 from the John Templeton Foundation.
AP thanks Aspen Center for Physics, which is supported by National Science Foundation grant PHY-2210452, for hospitality during the work on this project.
N\v{C}, HL, and  AP thank the
Simons Center for Geometry and Physics (Stony Brook) for its hospitality during
the program ’Black Hole Physics from Strongly Coupled Thermal Dynamics,’ where
part of this work was completed.

%%%%%%%%%%%%%%%%%%%%%%%%%%%%%%%%%%%%%%%%%%%%%%%%%%%%%
%%%%%%%%%%%%%%%%%%%%%%%%%%%%%%%%%%%%%%%%%%%%%%%%%%%%%

\appendix

%%%%%%%%%%%%%%%%%%%%%%%%%%%%%%%%%%%%%%%%%%%%%%%%%%%%%
%%%%%%%%%%%%%%%%%%%%%%%%%%%%%%%%%%%%%%%%%%%%%%%%%%%%%
\section{Spacelike geodesics in charged black brane}

In this appendix we investigate spacelike geodesics on the charged black brane background, partially following \cite{Brecher:2004gn}. First we show that, analogously to the neutral case, in the Euclidean regime, the large energy limit of the geodesic computation reconstructs order-by-order the T+J sector on the boundary. 

Then we look at a Lorentzian regime and consider imaginary charge. We show that in the large energy regime we precisely reproduce the prescription for the critical time $\tau_c(\mu,q)$ \eqref{eq:TauCQExp2} obtained in the main body of the paper from the null-geodesic analysis. 

\subsection{Spacelike geodesics in the Euclidean section} \label{app:SpaceGeo}

Let us start with the metric in Euclidean signature
\begin{align}
    \label{eq:ChargedMet1b}
    ds^2 = r^2\,f(r)\,d\tau^2 + \frac{dr^2}{r^2\,f(r)}+ r^2\,dx_{3}^2\qq{where}
    f(r) = 1- \frac{\mu}{r^4}+ \frac{q^2}{r^6}\,.
\end{align}
In the above, we have that the boundary spacetime (located at $r=\infty$) is $\mathbb{R}^{3}\times\mathbb{C}$ and we have set the radius of AdS $R_{\rm AdS}=1$. In general, one could work in arbitrary dimensions, but we shall focus only on the case with a five dimensional bulk. 

Let us focus on spacelike geodesics that have no spatial dependence. That is, let us look at spacelike geodesics by introducing conserved charges
\begin{align}
    \tilde{E}= r^2\,f(r)\,\dot\tau\,,\qquad P = r^2\,\dot x\,,
\end{align}
and set $P=0$. Then the condition that the geodesic is spacelike at each point reads
\begin{align}
    r^2\,f(r)\,\dot \tau^2 + \frac{\dot r^2}{r^2\,f(r)} = \frac{\tilde{E}^2+ \dot r^2}{r^2\,f(r)} =1\quad\Rightarrow\quad\dot r^2 = r^2 f(r) - \tilde{E}^2\,.
\end{align}
The turning point equation is then given by $r^2f(r) = \tilde{E}^2$, which, defining $y = r^2$, can be rewritten as
\begin{align}
    y^3- \tilde{E}^2\,y^2 - \mu \,y + q^2 \equiv (y-y_1)(y-y_2)(y-y_3) = 0\,. 
\end{align}
The largest positive root (and there will always be a largest positive root for real coefficients $\tilde{E}^2$, $\mu$, $q^2$) we denote by $y_1$, and will be the turning point. 
 
We will be first interested in the $\tilde{E}\to \infty$ limit of the correlator. Recall that if $q=0$, then one of the roots goes to 0 and the two roots of the equations are 
\begin{align}
    y^{\pm}_{q=0} = \frac12\left(\tilde{E}^2 \pm \sqrt{\tilde{E}^4 + 4\mu}\right)\,,
\end{align}
meaning that in the large $\tilde{E}$ limit, the leading order solutions scale as
\begin{align}
    y_{q=0}^{+} \approx \tilde{E}^2 + \frac{\mu}{\tilde{E}^2}+\ldots\,,\qquad y_{q=0}^{-} \approx - \frac{\mu}{\tilde{E}^2}+\ldots\,,
\end{align}
so these are the starting points for the turning points as one includes non-zero charge. One finds that the three turning points are given by 
{\allowdisplaybreaks{\small{
\begin{subequations}
    \begin{align}
        y_1 &\approx \tilde{E}^2 + \frac{\mu}{\tilde{E}^2}- \frac{q^2}{\tilde{E}^4} - \frac{\mu^2}{\tilde{E}^6} + \frac{3\,q^2\,\mu}{\tilde{E}^8}+\ldots\,,\\
   y_2 &\approx -\frac{q}{\tilde{E}} -\frac{\mu }{2
   \tilde{E}^2}-\frac{\mu ^2}{8 \tilde{E}^3 q}+\frac{q^2}{2
   \tilde{E}^4}+\frac{\mu ^4+96 \mu  q^4}{128 \tilde{E}^5 q^3}+\ldots \,.\\
        y_3 &\approx \frac{q}{\tilde{E}} -\frac{\mu }{2
   \tilde{E}^2}+\frac{\mu ^2}{8 \tilde{E}^3 q}+\frac{q^2}{2
   \tilde{E}^4}-\frac{\mu ^4+96 \mu  q^4}{128 \tilde{E}^5 q^3}+\ldots \,,
    \end{align}
\end{subequations}}}}%
where notice that $y_2$ and $y_3$ are solutions that are the same if $q\to -q$, or alternatively
\begin{align}
    y_2(-q) = y_3(q)\,.
\end{align}
Note that it might be naviely weird that when $q\to 0$, $y_2 =y_3$, but notice that the subleading terms in the large $\tilde{E}$ expansion diverge when $q\to 0$. This is the sign that the limits $q\to0$ and $\tilde{E}\to \infty$ do not commute. 

\paragraph{Proper time integral.}
We now want to analyse the integral
\begin{align}
    \tau = \int\frac{\dot \tau}{\dot r}\,dr = 2\int_{r_t}^{\infty}\frac{\tilde{E}\,dr}{r^2f(r)\sqrt{r^2f(r)-\tilde{E}^2}}\,,
\end{align}
where we have assumed that the geodesics are symmetric, thus getting twice the integral from the boundary to the turning point. 

Again using $y= r^2$ and that $r_t^2 = y_1$, we get
\begin{align}
    \tau = \tilde{E}\int_{y_1}^{\infty}\frac{dy\,y^{\frac52}}{(y^3-\mu y +q^2)\sqrt{(y-y_1)(y-y_2)(y-y_3)}
    }\,,
\end{align}. 
Following \cite{Brecher:2004gn}, we define
\begin{align}
\label{eq:wicubic}
    (y^3-\mu y +q^2) \equiv (y-w_1)(y-w_2)(y-w_3)\,.
\end{align}
From \eqref{eq:wicubic} we find the following relations:
\begin{align}
\label{eq:wrelations}
    w_1+w_2 + w_3 = 0\,,\qquad w_1\,w_2\,w_3 = -q^2\,, \qquad w_1\,w_2 + w_1\,w_3 + w_2\,w_3 = -\mu\,.
\end{align}
The integral that we then analyse is 
\begin{align}
    \tau = \tilde{E}\int_{y_1}^{\infty}\frac{dy\,y^{\frac52}}{(y-w_1)(y-w_2)(y-w_3)\sqrt{(y-y_1)(y-y_2)(y-y_3)}
    }\,.
    \label{eq:tauint0}
\end{align}
One can put this integral into the form
\begin{align}
    \tau &= \frac{2\tilde{E}\,y_1^{\frac52}}{\sqrt{y_3-y_2}(y_1-w_1)(y_1-w_2)(y_1-w_3))}\Bigg[\sum_{i=1}^{3}\frac{\hat c_i(c_i-s a)^3}{c_i^3} \Pi\left(\frac{c_i}{a}, \arcsin{\sqrt{a}}, s\right)\nonumber\\*
    &\quad + \,\frac{a^3\,s^3}{c_1\,c_2\,c_3}\,F\left(\arcsin{\sqrt{a}}, s\right)\Bigg]\,,
    \label{eq:TauResCharged}
\end{align}
where
\begin{align}
   a= \frac{y_3-y_2}{y_1-y_2}\,,\qquad  c_i \equiv \frac{y_3-w_i}{y_1-w_i}\,,\qquad s\equiv \frac{y_3}{y_1\,a} < \frac{1}{a}\,,
\end{align}
and
\begin{align}\label{e.tretiaztroch}
    \hat c_1 = \frac{c_1^2}{(c_1-c_2)(c_1-c_3)}\,,\qquad
    \hat c_2 = \frac{c_2^2}{(c_2-c_1)(c_2-c_3)}\,,\qquad
    \hat c_3 = \frac{c_3^2}{(c_3-c_1)(c_3-c_2)}\,.
\end{align}

\paragraph{Proper length integral.}
The regularised proper length is given by 
\begin{align}
    \label{eq:PropLenCharged}
    \cL = \lim_{\Lambda\to\infty}\left[2\int_{r_t}^{\Lambda}\frac{dr}{\sqrt{r^2\,f(r)-\tilde{E}^2}}-2\log\Lambda\right]\,,
\end{align}
where $\Lambda$ is the near-boundary cut-off. Let us focus on the integral
\begin{align}
    \cL_{\Lambda} \equiv 2\int_{r_t}^{\Lambda}\frac{dr}{\sqrt{r^2\,f(r)-E^2}} = \int_{y_1}^{\Lambda^2}\frac{dy\,\sqrt{y}}{\sqrt{(y-y_1)(y-y_2)(y-y_3)}}\,,
\end{align}
where we again introduced $r^2=y$. This integral can be written as
\begin{align}
    \cL_{\Lambda} 
    &= \frac{2\sqrt{y_1}}{\sqrt{y_3-y_2}}\Bigg[s\,a F(\arcsin\chi_{\Lambda},s)+(1-s\,a) \Pi\left(\frac{1}{a}, \arcsin\chi_{\Lambda},s\right) \Bigg]\,,
    \label{eq:UnregularisedPropLengt}
\end{align}
where
\begin{align}
    \frac{dy}{\sqrt{(y-y_1)(y-y_3)}} = \frac{\sqrt{a}\,dy}{(y-y_3)\,\chi}= \frac{2(y-y_3)}{\sqrt{a}(y_1-y_3)}\,d\chi\,.
\end{align}

To get the renormalized length, one needs to extract the logarithmic divergence from the above expression. Here we follow \cite{Brecher:2004gn}. The idea is to find a logarithmic branch point. These appear when the arguments of the first and second variable of the elliptic-$\Pi$ integral coincide. One finds that the function $\Pi(n,z,m)$ for fixed $n$ and $m$ has a logarithmic branch point when
\begin{align}
    z= \pm \arcsin\left(\frac{1}{\sqrt{n}}\right) + k\pi\,.
\end{align}
Note there are other branch points, but it is enough to examine the above class. In our case, $n = 1/a$, so as $\Lambda\to \infty$, $z \to \arcsin(\sqrt{a})$, which precisely gives us the logarithmic divergence. We can use the following functional identity
\begin{equation}\label{e.ebenso}
\begin{split}
    \Pi\left(\frac{1}{a}, \arcsin\chi_{\Lambda},s\right) &= F(\arcsin\chi_{\Lambda},s)- \Pi\left(s\,a, \arcsin\chi_{\Lambda},s\right)\\*
    &-\frac{1}{2\sqrt{\frac{(1-s a)(    1-a)}{a}}}\log\left[\frac{\sqrt{1-s\,\chi_{\Lambda}^2}- \sqrt{\frac{(1-s a)(1-a)}{a}}\,\frac{\chi_{\Lambda}}{\sqrt{1-\chi_{\Lambda}^2}}}{\sqrt{1-s\,\chi_{\Lambda}^2}+\sqrt{\frac{(1-s a)(1-a)}{a}}\,\frac{\chi_{\Lambda}}{\sqrt{1-\chi_{\Lambda}^2}}}\right]\,.
\end{split}
\end{equation}
Note that for the choice where $y_1>y_3>y_2$, $a>0$, $(1-sa)>0$ and $1-a>0$.

To get the expansion at leading order in $\Lambda$ we expand
\begin{align}
    \chi_{\Lambda} = \sqrt{a}+ \frac{\sqrt{a}(y_3-y_1)}{2\,\Lambda^2}+\ldots\,,
\end{align}
Expanding the argument of the logarithm in \eqref{e.ebenso} we find
\begin{align}
    \frac{\sqrt{1-s\,\chi_{\Lambda}^2}- \sqrt{\frac{(1-s a)(1-a)}{a}}\,\frac{\chi_{\Lambda}}{\sqrt{1-\chi_{\Lambda}^2}}}{\sqrt{1-s\,\chi_{\Lambda}^2}+\sqrt{\frac{(1-s a)(1-a)}{a}}\,\frac{\chi_{\Lambda}}{\sqrt{1-\chi_{\Lambda}^2}}} \approx \frac{y_1-y_2+y_3}{4\,\Lambda^2}\left(1+ \coo{\Lambda^{-2}}\right)\,,
\end{align}
When the logarithm of this is taken, with $\Lambda\to\infty$, one just obtains  
\begin{align}
    \log\left(\frac{y_1-y_2+y_3}{4}\right)- 2\log\Lambda\,.
\end{align}
Note that because $y_2<0$, and $y_1>y_3>0$, we are summing over three positive terms, so the above expression should be well defined. %
Overall we get
\begin{align}
    \cL_{\Lambda} 
    &= \frac{2\sqrt{y_1}}{\sqrt{y_3-y_2}}\Bigg[ F(\arcsin\sqrt{a},s)-(1-s\,a)\, \Pi\left(s a , \arcsin\sqrt{a},s\right) \Bigg]-  \log\left(\frac{y_1-y_2+y_3}{4}\right)\nonumber\\*
    &\quad + 2\log\Lambda+ \coo{\Lambda^{-2}}\,,
    \label{eq:UnregularisedPropLength2}
\end{align}
so then, when one completely regularises the proper length, \eqref{eq:PropLenCharged}, the result reads
\begin{align}
    \label{eq:PropLengthChargedResults}
    \cL
    &= \frac{2\sqrt{y_1}}{\sqrt{y_3-y_2}}\Bigg[ F(\arcsin\sqrt{a},s)-(1-s\,a)\, \Pi\left(s a , \arcsin\sqrt{a},s\right) \Bigg]-  \log\left(\frac{y_1-y_2+y_3}{4}\right)\,.
\end{align}

Let us now take the result for the proper time \eqref{eq:TauResCharged}-\eqref{e.tretiaztroch} and expand each term individually in large $\tilde{E}$, keeping $q$ and $\mu$ fixed.

Start by expanding the $\arcsin\sqrt{a}$, that goes to 0 as 
\begin{align}
    \arcsin\sqrt{a} \approx \frac{\sqrt{2q}}{E^{\frac32}}+\ldots\,,
\end{align}
so one can expand the middle argument of $\Pi$ and the first argument of $F$ around 0. On the other hand, in the large $\tilde{E}$ limit $s$ goes to 1/2
\begin{align}\label{e.suse}
    s \approx \frac12 - \frac{\mu}{4q\,\tilde{E}}+ \ldots\,.
\end{align}
Finally, all first arguments actually blow up
\begin{align}
    \frac{c_i}{a} \approx -\frac{w_i\,\tilde{E}}{2\,q}+ \frac{1}2+\ldots\,.
\end{align}
To consistently take the large $\tilde{E}$ limit, we first expand both elliptic functions in $\arcsin\sqrt{a}$, then insert $s$, expand in large $\tilde{E}$ and finally for the Elliptic-$\Pi$ functions also insert the values for $c_i/a$. What we get in the end is the following expansion
\begin{align}
\label{eq:TauESeries}
    \tau \approx \frac{2}{\tilde{E}} -\frac{8\mu}{5\tilde{E}^5} + \frac{16q^2}{7\tilde{E}^7}+\frac{32\mu^2}{9\tilde{E}^9}- \frac{128q^2\,\mu}{11\tilde{E}^{11}}+\frac{128 \left(6 q^4-7 \mu ^3\right)}{91 \tilde{E}^{13}}+ \frac{256q^2\mu^2}{5\tilde{E}^{15}}\ldots\,.
\end{align}
This can be inverted as 
\begin{align}\label{e.inversionjk}
    \tilde{E}\approx \frac{2}{\tau}-\frac{\mu}{10}\tau^3+ \frac{q^2}{28}\tau^5-\frac{11\mu^2}{1800}\tau^7+\frac{q^2\mu}{154}\tau^9+\frac{-4361 \mu ^3-13500 q^4}{7644000}\tau^{11}+\ldots 
\end{align}

Next we need to expand the proper length, which is given by Equation\ \eqref{eq:PropLengthChargedResults}.
We follow an analogous approach as for the time integral. The final result works out to be
\begin{align}
\label{eq:cLExpansion}
    \cL \approx 2\log 2 - 2\log \tilde{E} -\frac{2\mu}{\tilde{E}^4} +\frac{8q^2}{3\tilde{E}^6} + \frac{4\mu^2}{\tilde{E}^8} - \frac{64q^2\,\mu}{5\tilde{E}^{10}} +\frac{32(6q^4-7\mu^3)}{21\tilde{E}^{12}}+\ldots  
\end{align}
Using now the expression $\tilde{E}(\tau)$ \eqref{e.inversionjk}, we find
\begin{align}
    \cL = 2\log\tau-\frac{\mu\tau^4}{40} +\frac{q^2\tau^6}{168} -\frac{11\mu^2\tau^8}{14400}+\frac{q^2\mu\tau^{10}}{1540}+\frac{-4361 \mu ^3-13500 q^4}{91728000}\,\tau^{12}+\ldots\,.
    \label{eq:GeoLength}
\end{align}

\paragraph{Proper length from the OPE.}

We want to compare the above geodesic results with the stress-charge spectrum of the correlator computed holographically in Section \ref{ssec:qneq0ansatz}. Let us remind that the boundary thermal correlator can be decomposed in conformal blocks
\begin{align}
    G_{\rm T+J}(\tau) = \frac{1}{\tau^{2\Delta}}\left(1+\sum_{n=0}^{\infty}\Lambda_n(q)\,\tau^{4+2n}\right)\,,
\end{align}
where the coefficients $\Lambda_n(q)$ can be computed to high order holographically using the method described in Appendix \ref{app:OPE}.

To compare this with the proper length of the geodesic, we consider
\begin{align}
    L\equiv - \lim_{\Delta\to\infty} \frac{1}{\Delta}\log G(\tau)\,,
\end{align}
which gives
\begin{align}
    L = 2 \log \tau - \frac{\mu\tau^4}{40} +\frac{q^2\tau^6}{168} -\frac{11\mu^2\tau^8}{14400}+\frac{q^2\mu\tau^{10}}{1540}+\left(-\frac{89}{1872000}\mu ^3+\frac{3}{20384} q^4\right)\tau^{12}+\ldots\,.
    \label{eq:CFLenght}
\end{align}
The two expressions \eqref{eq:CFLenght} and \eqref{eq:GeoLength} agree up to all examined orders in $\tau$.

We conclude that, as in the neutral case, in the small $\tau$-regime the geodesic analysis for the \textit{whole} correlator precisely reproduces the T+J sector. This provides strong evidence that the map between different sectors of the boundary correlator and bulk geodesics proposed in \cite{Ceplak:2024bja} still holds, with the stress tensor naturally generalized to the T+J  sector.

\subsection{Time shift of a bouncing geodesic in a spacetime with imaginary charge.}

Let us now consider spacelike geodesics in the Lorentzian section of the charged black hole background with imaginary charge $q=-i\tilde{q}$ with $\tilde{q}\in\mathbb{R}$.
In this case, one expects an ``ordinary'' bouncing geodesic in the limit of large (Lorentzian) energy.

To show this, we compute the time-shift of spacelike geodesics with energy $E$
\begin{align}
    \tau = iE\int_{y_1}^{\infty}\frac{dy\,y^{\frac52}}{(y-w_1)(y-w_2)(y-w_3)\sqrt{(y-y_1)(y-y_2)(y-y_3)}
    }\ ,
\end{align}
where $y_1$, $y_2$ and $y_3$ are the solutions of the equation
    \begin{equation}
    y^3+E^2y^2-y-\tilde{q}^2=0\ .
    \end{equation}
One can express this integral as a linear combination of Elliptic integrals following the same calculation as presented in the previous subsection.
Then one takes the limit $E\to \infty$ followed by the small $\tilde q$ expansion. 
At leading order in the large energy expansion, we find
    \begin{equation}\label{e.tauzebri}
    \tau=-\frac{2i}{E}+\left(\frac{1}{2}+\frac{i}{2}\right)\frac{\pi}{\mu^{\frac14}} +\left(-\frac{3}{8}+\frac{3 i}{8}\right) \frac{\pi\,\tilde{q}^{\phantom{.}2}}{\mu^{\frac74}} +\frac{i\pi\,(\tilde{q}^{\phantom{.}2})^{\frac32}}{\mu^{\frac52}}+\left(\frac{45}{64}+\frac{45 i}{64}\right) \frac{\pi\,\tilde{q}^{\phantom{.}4}}{\mu^{\frac{13}{4}}}+\ldots\ ,
    \end{equation}
where the ellipsis represent terms of higher order in $\tilde{q}$ and subleading terms in the $E\rightarrow\infty$ limit. 
Inverting this expression we get the Lorentzian energy $E$ as a function of the boundary time $\tau$.
The proper length in the strict $E\to \infty$ limit is given by  
    \begin{equation}
    \mathcal{L}\approx2\log(\tau-\tau_c(\mu,\tilde q))\ ,
    \end{equation}
where the critical time $\tau_c(\mu,\qt)$ can be read off from \eqref{e.tauzebri}.

To get the time-shift of a null geodesic in the spacetime where the charge is real, we take the critical point of the bouncing geodesic in the spacetime with imaginary charge and analytically continued to $\tilde{q}=iq$
    \begin{equation}\label{e.finofaAa}
    \begin{split}
    \tau_c(\mu,q)=&\left(\frac{1}{2}+\frac{i}{2}\right) \frac{\pi}{\mu^{\frac14}} +\left(\frac{3}{8}-\frac{3 i}{8}\right) \frac{\pi\,q^2}{\mu^{\frac74}} -\frac{\pi q^3}{\mu^{\frac52}}+\left(\frac{45}{64}+\frac{45 i}{64}\right)\frac{\pi\,q^4}{\mu^{\frac{13}{4}}}+\ldots\\
    =&\,\tau_c(\mu,0) + \frac{6\,q^2}{\pi^6}\tau_c(\mu,0)^7+ \frac{32\,i\,q^3}{\pi^9}\tau_c(\mu,0)^{10} - \frac{90\,q^4}{\pi^{12}}\tau_c(\mu,0)^{13} + \ldots\ ,
    \end{split}
    \end{equation}
where in the second line we used 
    \begin{equation}
    \tau_c(\mu,0)\equiv\left(\frac{1}{2}+\frac{i}{2}\right) \frac{\pi}{\mu^{\frac14}}\ .
    \end{equation}
    This is \eqref{eq:TauCQExp2}, the time-shift obtained from the contour depicted in Figure~\ref{fig:4dContStandard} in the black brane with real charge.

\subsection{Null geodesics as a limit of complex spacelike geodesics}\label{app:bounce}

For a black hole with imaginary charge, we can find real bouncing geodesic for all values of the imaginary charge $\qt$.
In addition, these bouncing geodesics are the infinite energy limit of real spacelike geodesics connecting points on the opposite boundaries of AdS, as depicted in Figure~\ref{fig:Penrose}. 
This picture is completely analogous to the one in neutral black holes. 
In the spacetime with real charge the real bouncing geodesic that reflects off the timelike singularity is the high-energy limit of \emph{complex} spacelike geodesics.%
\footnote{Complex spacelike geodesics already play a crucial role in correlation functions on neutral black brane backgrounds where they correspond to the saddles to contribute to the two-sided correlator in the semiclassical limit \cite{Fidkowski:2003nf, Festuccia:2005pi}. They also play a crucial role in de Sitter correlation functions \cite{Chapman:2022mqd,Aalsma:2022eru} and in the analysis of holographic timelike entanglement entropy \cite{Heller:2024whi, Heller:2025kvp}.}
The spacelike geodesics have to be complex, since they are connecting points located on timelike separated boundaries.

\begin{figure}[t]
    \centering
    \includegraphics[width=\linewidth]{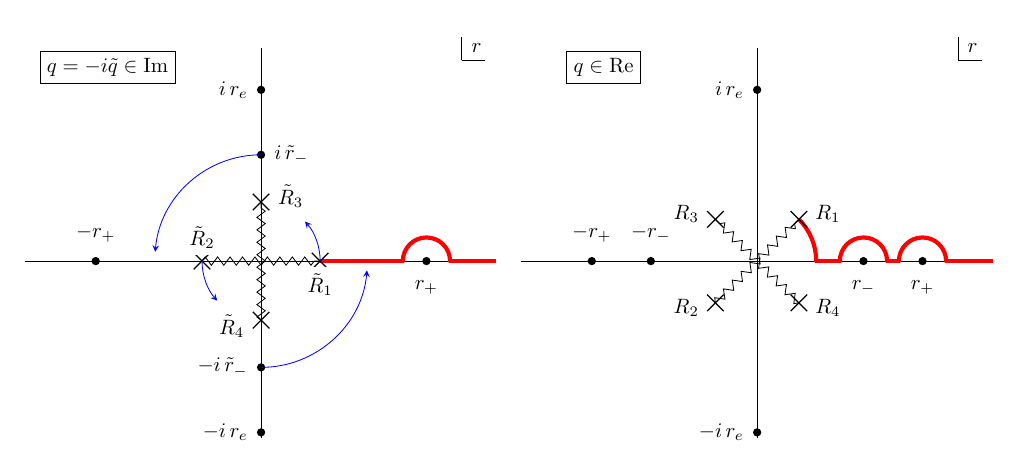}
    \caption{Left: The complex contour associated with spacelike geodesics in the black brane spacetime with imaginary charge. The crosses denote $\tilde R_i$, the solutions to the turning point equation \eqref{eq:TurningPoint}  (with $\tilde R_{5,6}$ not pictured).  These are as branch points in the computation of the time-shift $\tau_c$. When the charge is analytically continued to real values, the location of the inner horizon becomes real while the turning point becomes complex. On the right, we show the contour for the complex spacelike geodesic in the spacetime with real charge. For large energies, the spacelike geodesic closely follows the real null geodesic everywhere apart from very close to the singularity, where the turning point becomes complex.}
    \label{fig:ComplexR}
\end{figure}

Start by considering spacelike geodesics connecting points on opposite boundaries of the AdS in the spacetime with imaginary charge. 
Such  geodesics probe the spacetime behind the horizon: their turning point is determined by 
\begin{align}
    E^2 = - r^2 f(r)\,,
\end{align}
which can be rewritten as 
\begin{align}
\label{eq:TurningPoint}
    r^6 + E^2 \,r^4 - \mu\,r^2 - \qt^2= 0\,.
\end{align}
In our discussion we will only focus on the high-energy limit since this is the most important regime for the analysis of bouncing geodesics.
As $E \to \infty$, the six roots of this equation are to leading order in $1/E$ expansion.
\begin{align}
\label{eq:TurningPoints}
    \tilde R_{1,2} = \pm \sqrt{\frac{\qt}{E}}\,, \qquad \tilde R_{3,4} = \pm i \sqrt{\frac{\qt}{E}}\,, \qquad \tilde R_{5,6} = \pm i E\,.
\end{align}
The turning point is the largest positive root, $\tilde R_1$, which vanishes as $E\to \infty$.
In order to calculate the appropriate time-shift of the geodesic, one needs to specify the contour along which one avoids the inner horizon \cite{Festuccia:2005pi}, depicted in Figure~\ref{fig:ComplexR}.%
\footnote{In this discussion we only focus on spacelike geodesics that probe the region close to the future singularity. Geodesics that probe the spacetime close to the past singularity are associated with a different choice of contour \cite{Festuccia:2005pi}.}
As the energy is increased, the turning point goes closer and closer to the singularity, until in the limit $E\to \infty$ it becomes the bouncing geodesic that reaches the singularity. 

Let us briefly discuss the properties of the spacelike geodesics near the singularity. 
From the definition of the energy of the geodesic, we can find
\begin{align}
    \dot t = \frac{E}{r^2f(r)} \xrightarrow[r\to \tilde R_1]{} -\frac{\tilde R_1^4\,E}{\tilde q^2} = -\frac{1}{E} <0 \,,
\end{align}
where the dot denotes the derivative with respect to the parameter parametrising the geodesic and we approximated the denominator with the term that dominates near the origin. 
We see that the time decreases near the turning point, which is appropriate given the direction of time near the future singularity. 
Similarly, one can use \eqref{eq:PotentialIntro} to find
\begin{align}
    \ddot r = r\left(1+ \frac{\mu}{r^4} + \frac{2\tilde q^2}{r^6}\right) \xrightarrow[r\to \tilde R_1]{} \frac{2\qt^2}{\tilde R_1^5} >0\,,
\end{align}
meaning that the turning point is indeed a minimum distance to the origin.
Using the time-reflection symmetry of the geometry one can then to extend the geodesic across the $t = 0$ surface and thus connect points on opposite asymptotic boundaries using spacelike geodesics.

Let us now analytically continue these spacelike geodesics to real charge $\tilde q = i q$. 
From \eqref{eq:RootExpansions} we see that $r_+$ and $i r_e$ stay real and imaginary respectively, while $\tilde r_-$ rotates by $\pi/2$ in the anti-clockwise direction to $r_-$. 
On the other hand, the turning points $\tilde R_{1,2,3,4}$ all rotate only by $\pi/4$, as depicted in Figure~\ref{fig:ComplexR} so that
\begin{align}
    R_{1,2,3,4} = e^{\frac{i\pi}{4}}\tilde R_{1,2,3,4}\,,\qquad R_{5,6} = \tilde R_{5,6}\,,
\end{align}
That means that in the spacetime with real charge, the turning point of spacelike geodesics that limit to the bouncing geodesic is complex.

The contour of integration has to be deformed in such a way that no singularities cross it during analytic continuation. 
This then gives us a natural prescription for the contour that we employ for the null case in Figure~\ref{fig:4dContStandard}.
The most interesting behaviour arises near the turning point, where the contour needs to be deformed away from the real line to end on the complex turning point.

Let us examine this region in a bit more detail. 
Let us consider a highly energetic spacelike geodesic in the spacetime with real charge. 
The condition
\begin{align}
    E^2 = \dot r^2 - r^2 f(r)\,,
\end{align}
tells us that for all $r$ such that $E^2 \gg |r^2 f(r)|$, the spacelike geodesic will closely approximate the null bouncing geodesic, which is depicted on the Penrose diagram in Figure~\ref{fig:PDRealQ}.
Significant deviations will arise 
only very close to the singularity where   $E^2 \approx |r^2 f(r)|$.

Let us now track what happens with the time and radial directions as the spacelike geodesic traverses the geometry.
The change of time along the path is characterised by
\begin{align}
    \dot t  = \frac{E}{r^2\,f(r)}\,,
\end{align}
and we can use approximate this rate using the trajectory of the null geodesic: In region I, $\dot t >0$, it switches sign in region II, and it positive again in region VII. 
Very close to the singularity, the spacelike geodesic starts to deviate significantly and become complex.
In fact, near the complex turning point $R_1$, the time rate is negative
\begin{align}
    \dot t  = \frac{E}{r^2\,f(r)} \xrightarrow[r \to R_1]{} -\frac{1}{E}<0\,.
\end{align}
Similarly, one can analyse the second derivative of the change in the radial direction. 
For the real charge, we find
\begin{align}
    \ddot r = r\left(1+ \frac{\mu}{r^4} - \frac{2q^2}{r^6}\right) \xrightarrow[r\to R_1]{} -\frac{2 q^2}{R_1^5} = 2 \,e^{-\frac{i\pi}{4}}\,\frac{E^{\frac52}}{q^{\frac12}} \,,
\end{align}
most importantly
\begin{align}
    \mathrm{Re}(\ddot r) >0\,, \qquad \mathrm{Im}(\ddot r) <0\,.
\end{align}
meaning that at the turning point, the real component of the radial direction is at its minimum value while the imaginary component is at its maximum value. 

All in all, this paints the following picture of spacelike geodesics at high energies in the spacetime with real charge. 
Far away from the singularity, they closely follow the null bouncing geodesic, from which they deviate only near the singularity. There they escape the real section of the geometry and acquire a small imaginary component in the radial direction.
Using again the time-reflection symmetry of the geometry, one can again continue these geodesics across the $t=0$ surface in the region VII of the Penrose diagram in Figure~\ref{fig:PDRealQ}, obtaining (complex) spacelike geodesics that connect points on different copies of the asymptotic boundary on the same side of the Penrose diagram.

%%%%%%%%%%%%%%%%%%%%%%%%%%%%%%%%%%%%%%%%%%%%%%%%%%%%%
%%%%%%%%%%%%%%%%%%%%%%%%%%%%%%%%%%%%%%%%%%%%%%%%%%%%%

\section{Holographic extraction of OPE coefficients}
\label{app:OPE}

In this Appendix we explain the holographic computation for the extraction of the CFT data for the stress-charge sector of the boundary correlator. The method  is based on a holographic ansatz introduced in \cite{Fitzpatrick:2019zqz}, that provides an effective way to extract OPE data in a large class of holographic CFTs.\footnote{See \cite{Li:2019tpf,Fitzpatrick:2020yjb,Karlsson:2022osn,Huang:2022vet,Esper:2023jeq} for generalisations to other bulk theories as well as to external operators with spin.}  Here we provide a generalization that allows us to extract data from the whole stress-charge sector.

    \subsection{Holographic ansatz}

The aim is to solve the bulk equation of motion 
    \begin{equation}\label{eq:weworig}
        (\Box-m^2)\phi=0\, ,\qquad m^2=\Delta(\Delta-d)
    \end{equation}
in the background
    \begin{equation}\label{e.znovachbb}
    ds^2 = r^2f(r)d\tau^2 + \frac{dr^2}{r^2f(r)}+ r^2d\vec{x}_{d-1}^{\phantom{.}2}\qq{where}
    f(r) = 1- \frac{\mu}{r^{d}}+ \frac{q^2}{r^{2d-2}}\,,
    \end{equation}
in a way that mirrors the OPE structure of the correlator \eqref{eq:OPEres1}. To do this, first, transform coordinates $(\tau,\vec{x},r)\rightarrow(w,\rho,r)$
    \begin{equation}
    \rho^2=r^2\,\vec{x}^{\phantom{.}2}\,,\qquad w^2=1+r^2(\tau^2+\vec{x}^{\phantom{.}2})    \,,
    \end{equation}
and assume
    \begin{equation}\label{eq:anz1}
        \phi(w,\rho,r)=\left(\frac{r}{w^2}\right)^\Delta \psi(w,\rho,r)\ ,
    \end{equation}
where $(r/w^2)^\Delta$ is the solution in the pure AdS space. In this language, the boundary correlator can be obtained by taking the limit
    \begin{equation}\label{eq:LimGen}
        G(\tau,\vec{x})
        =\frac{1}{(\tau^2+\vec{x}^{\phantom{.}2})^\Delta}\lim_{r\rightarrow\infty}\psi\,.
    \end{equation}
In holographic CFTs one expects two sectors: The stress-charge sector and the double-trace sector. In the bulk these correspond to rewriting
    \begin{equation}
    \psi=\psi_{\rm T+J}+\psi_{[\phi\phi]}\ ,
    \end{equation}
where $\psi_{\rm T+J}$ corresponds to the contributions from the identity operator, stress-tensor, electric current and their composites, while $\psi_{[\phi\phi]}$ is the contribution due to the double-trace operators. The former is fully determined by the expansion near the conformal boundary, while the double-trace sector requires information about the interior of the bulk geometry and is not captured by this method \cite{Fitzpatrick:2019zqz,Karlsson:2022osn}.

\begin{table}[!hb]
\centering
\setlength{\tabcolsep}{16pt}
\renewcommand{\arraystretch}{0.92}
\begin{tabular}{@{}c|c@{}}
\hline
\textbf{{\small{$(a,\,b,\,c)$}}} & \textbf{{\small{$C^{\,i,j,k}_{a,b,c}$}}} \\
\hline
{\small{(0,\,0,\,2)}} & {\footnotesize{$2j(1+2j)$}} \\
{\small{(0,\,1,\,1)}} & {\footnotesize{$4 (2 i - k) (k - \Delta)$}} \\
{\small{(0,\,1,\,2)}} & {\footnotesize{$4 (i - j - k) (-2 + i - j - k + \Delta)$}} \\
\hline
{\small{(2,\,0,\,2)}} & {\footnotesize{$-2 j (1 + 2 j) \mu$}} \\
{\small{(2,\,1,\,0)}} & {\footnotesize{$-8 (-1 + k - \Delta) (k - \Delta) \mu$}} \\
{\small{(2,\,1,\,1)}} & {\footnotesize{$-2 (k - \Delta) (3 + 8 i - 4 j - 8 k + 
   4 \Delta) \mu$}} \\
{\small{(2,\,1,\,2)}} & {\footnotesize{$-((8 i^2 + 8 (j + k) (1 + j + k) - 
     8 i (1 + 2 j + 2 k - \Delta) - 
     8 (j + k) \Delta + \Delta^2) \mu)$}} \\
{\small{(2,\,2,\,0)}} & {\footnotesize{$-4 (-1 + k - \Delta) (k - \Delta) \mu$}} \\
\hline
{\small{(3,\,0,\,2)}} & {\footnotesize{$2 j (1 + 2 j) q^2$}} \\
{\small{(3,\,1,\,0)}} & {\footnotesize{$8 q^2 (-1 + k - \Delta) (k - \Delta)$}} \\
{\small{(3,\,1,\,1)}} & {\footnotesize{$2 q^2 (k - \Delta) (5 + 8 i - 4 j - 8 k + 4 \Delta)$}} \\
{\small{(3,\,1,\,2)}} & {\footnotesize{$q^2 (8 i^2 + 4 (j + k) + 8 (j + k)^2 - 
   4 i (1 + 4 j + 4 k - 2 \Delta) + 2 \Delta - 
   8 (j + k) \Delta + \Delta^2)$}} \\
{\small{(3,\,2,\,0)}} & {\footnotesize{$4 q^2 (-1 + k - \Delta) (k - \Delta)$}} \\
\hline
{\small{(4,\,1,\,0)}} & {\footnotesize{$4 (-1 + k - \Delta) (k - \Delta) \mu^2$}} \\
{\small{(4,\,1,\,1)}} & {\footnotesize{$-4 (k - \Delta) (-1 - 2 i + 2 j + 
   2 k - \Delta) \mu^2$}} \\
{\small{(4,\,1,\,2)}} & {\footnotesize{$(2 i - 2 (j + k) + \Delta)^2 \mu^2$}} \\
\hline
{\small{(5,\,1,\,0)}} & {\footnotesize{$-8 q^2 (-1 + k - \Delta) (k - \Delta) \mu$}} \\
{\small{(5,\,1,\,1)}} & {\footnotesize{$4 q^2 (-3 - 4 i + 4 j + 4 k - 
   2 \Delta) (k - \Delta) \mu$}} \\
{\small{(5,\,1,\,2)}} & {\footnotesize{$-2 q^2 (1 + 2 i - 2 j - 2 k + \Delta) (2 i - 2 (j + k) + \Delta) \mu$}} \\
\hline
{\small{(6,\,1,\,0)}} & {\footnotesize{$4 q^4 (-1 + k - \Delta) (k - \Delta)$}} \\
{\small{(6,\,1,\,1)}} & {\footnotesize{$-4 q^4 (k - \Delta) (-2 i + 2 (-1 + j + k) - \Delta)$}} \\
{\small{(6,\,1,\,2)}} & {\footnotesize{$q^4 (2 i - 2 (-1 + j + k) + \Delta) (2 i - 
   2 (j + k) + \Delta)$}} \\
\hline
\end{tabular}
\caption{All non-zero elements of tensor $C_{a,b,c}^{i,j,k}$.}
\label{tab.cabctab}
\end{table}

In $d=4$, we assume the following near-boundary expansion for the T+J sector%
\footnote{The generalisation to other (even) dimensions is straightforward.}
    \begin{equation}
    \label{eq:theansatz}
        \psi_{\rm T+J}=1+\sum_{i=1}^\infty\,\sum_{j=0}^{\lfloor\frac{i}{2}\rfloor}\sum_{k=-\lfloor\frac{i}{2}\rfloor}^{2\lceil\frac{i}{2}\rceil-j}a_{j,k}^i\frac{\rho^{2j}w^{2k}}{r^{2\,i}}\ ,
    \end{equation}
where 1 corresponds to the contribution of the identity operator and the sum over $i$ contains the contributions from the stress-charge sectors according to the Table \ref{tab:t2}. Inserting this ansatz into the bulk equation of motion and expanding in $1/r$, one can determine the coefficients $a_{j,k}^i$ to arbitrary order in $i$. Most importantly, through the dictionary \eqref{eq:LimGen}, this large-$r$ expansion on the bulk side systematically maps to the OPE on the boundary, which for $\vec{x}=0$ is given by \eqref{eq:OPEres1}, resp.\ \eqref{eq:GenForm}.\footnote{Setting $\vec{x}=0$ corresponds to summing over different exchanged operators with different spins $J'$ but the same conformal dimension $\Delta'$, see Tab.\ \ref{tab:t2}.}
This way one extracts the OPE data $\Lambda_n(\mu,q)$; the first few coefficients are listed in \eqref{e.Lb00st}. Note that the powers of $\mu$ and $q$ count the number of stress-tensors and charge current contributions in the OPE, see Tab.\ \ref{tab:t2}. 
For certain scaling dimensions there will be several contributions from different exchanged operators. 
For example, in $\Lambda_5(\mu,q)$, corresponding to the exchange of operators with scaling dimension $\Delta'=12$, we find

{\footnotesize{
    \begin{equation}
    \begin{split}
    \Lambda_5(q)=&\frac{-639360 \Delta \!-\!1278432 \Delta ^2\!-\!1093216 \Delta ^3\!-\!93544 \Delta ^4\!-\!310750 \Delta ^5\!+\!419727 \Delta ^6\!-\!114114 \Delta ^7\!+\!9009 \Delta ^8}{3459456000 (\Delta -6) (\Delta -5) (\Delta -4) (\Delta -3) (\Delta -2)}\,\mu^3 \\
    &+ \frac{259164 \Delta \!+\!567759 \Delta ^2\!+\!405423 \Delta ^3\!+\!91603 \Delta ^4\!-\!25575 \Delta ^5\!-\!21780 \Delta ^6\!+\!3575 \Delta ^7}{201801600 (\Delta -6) (\Delta -5) (\Delta -4) (\Delta -3) (\Delta -2)}\,q^4\ ,
    \end{split}
    \end{equation}}}%
    with two terms coming from the exchange of $T^3$ and $J^4$ operators, whose contributions are proportional to  $\mu^3$ and $q^4$ respectively. This demonstrates how the simple power-counting of the holographic results helps us sort the individual contributions. 

\subsection{Recurrence relations}

The above method for computing the stress-charge correlator can be further improved by reformulating the problem in terms of the recurrence relations -- see \cite{Buric:2025fye} for more details.
Here we present the generalisation of this recurrence relation to the case of the charged black brane \eqref{e.znovachbb} in $d=4$.

Start by setting
    \begin{equation}
    a^i_{j,k}=
    \begin{cases}
    1&\text{for}\quad(i,j,k)=(0,0,0)  \\
    0&\forall\,(i,j,k)\notin I\cup\{(0,0,0)\}\, ,
    \end{cases}
    \end{equation}
where the index set $I=\{\,(i,j,k)\mid i\geq2\wedge0\leq j\leq\lfloor\frac{i}{2}\rfloor\wedge-\lceil\frac{i}{2}\rceil\leq k\leq2\lceil\frac{i}{2}\rceil-j\}$. To determine the remaining values of $a^i_{j,k}$ use the following algorithm:
    \begin{itemize}
        \item For each $i\geq2$
        \item For each $k\in\left[-\lceil\frac{i}{2}\rceil,\,2\lceil\frac{i}{2}\rceil\right]$
        \item For each $j\in\left[0,\,\lfloor\frac{i}{2}\rfloor\right]$
        \item Compute $a^i_{j,k}$ using
    \end{itemize}
    \begin{equation}
            a^i_{j,k}=-\frac{1}{C_{0,1,1}^{\,i,j,k}}\sum_{(a,b,c)\in J\backslash\{(0,0,0)\}}a^{i-a}_{j-b,\,k-c}\cdot C^{\,i-a,\,j-b,\,k-c}_{a,\,b+1,\,c+1}\ ,
    \end{equation}
where the index set $J=\{(a,b,c)\mid a\in[0,6]\wedge b\in[-1,1]\wedge c\in[-1,1]\}$ and the $C_{a,b,c}^{i,j,k}$ is a $7\times3\times3$ dimensional tensor for any fixed $i,\,j,\,k$, whose nonzero elements are listed in Table \ref{tab.cabctab}. The CFT data $\Lambda_n(q)$ can be obtained as
    \begin{equation}
    \Lambda_n(q)\equiv a^{n+2}_{0,\,n+2}\ .
    \end{equation}
This recursion relation method allows for extraction of the coefficients $\Lambda_n(q)$ up to order $n_{\max}\sim200-300$ on a standard laptop.

% %%%%%%%%%%%%%%%%%%%%%%%%%%%%%%%%%%%%%%%%%%%%%%%%%%%%%
% %%%%%%%%%%%%%%%%%%%%%%%%%%%%%%%%%%%%%%%%%%%%%%%%%%%%%

\bibliographystyle{JHEP}
\bibliography{draft} 

\end{document}